\documentclass[acmlarge,screen,nonacm]{acmart}
\graphicspath{{figures/}}
\usepackage{float}
\usepackage{placeins}

\usepackage[yyyymmdd]{datetime}

\usepackage{tabularx}
\usepackage{booktabs}
\usepackage{longtable}
\usepackage{dcolumn}
\newcolumntype{d}[1]{D{.}{.}{\#1}}

\usepackage{tikz}
\usetikzlibrary{arrows.meta, positioning, shapes.geometric, decorations.pathmorphing, calc}

\newcommand{\condnoai}{\textsc{Control}} %
\newcommand{\condai}{\textsc{AI-supported}} %
\newcommand{\condcomp}{AI-alone benchmark} %

\AtBeginDocument{%
 }

\setcopyright{none}
\acmDOI{}
\acmISBN{}
\acmConference{}{}{}
\acmBooktitle{}
\date{September 11, 2026}

\begin{document}

\title[When the AI Leaves the Tailorshop]{When the AI Leaves the Tailorshop: Measuring What an LLM Advisor Leaves Behind in Complex
Problem Solving}

\settopmatter{authorsperrow=3}

\author{Robin Welsch}
\orcid{0000-0002-7255-7890}
\affiliation{\institution{Aalto University}\city{Espoo}\country{Finland}}
\email{robin.welsch@aalto.fi}
\renewcommand{\shortauthors}{Robin Welsch}

\definecolor{cpsnoai}{HTML}{4477AA}
\definecolor{cpsai}{HTML}{EE6677}

\begin{abstract}

Complex problem solving depends on acting effectively and understanding how a system works. AI advice may support these outcomes unequally. Two preregistered experiments compared participants managing a simulated clothing factory with and without an LLM advisor. Across studies, AI-supported participants reported greater confidence and understanding with less effort. In the first study (N=200), assistance increased company value but produced no detectable prediction-accuracy difference. After withdrawal, previously supported participants outperformed controls when decisions were scored against repeating previous choices, but not default settings. Within the AI-supported group, more frequent recommendation alterations predicted better unaided performance. In the second study (N=198), AI-supported participants went bankrupt less often and showed a small knowledge advantage in the registered analysis, largely associated with remaining solvent. More frequent recommendation alterations predicted higher knowledge within the AI-supported group. Applied HAI evaluation should assess users’ understanding and independent capability alongside the performance achieved with AI support.
\end{abstract}

\begin{CCSXML}
<ccs2012>
 <concept>
 <concept\_id>10003120.10003121.10011748</concept\_id>
 <concept\_desc>Human-centered computing~Human computer interaction (HCI)~Empirical studies in HCI</concept\_desc>
 <concept\_significance>500</concept\_significance>
 </concept>
 <concept>
 <concept\_id>10010147.10010178</concept\_id>
 <concept\_desc>Computing methodologies~Artificial intelligence</concept\_desc>
 <concept\_significance>300</concept\_significance>
 </concept>
</ccs2012>
\end{CCSXML}


\begin{teaserfigure}
\centering
\includegraphics[width=\textwidth]{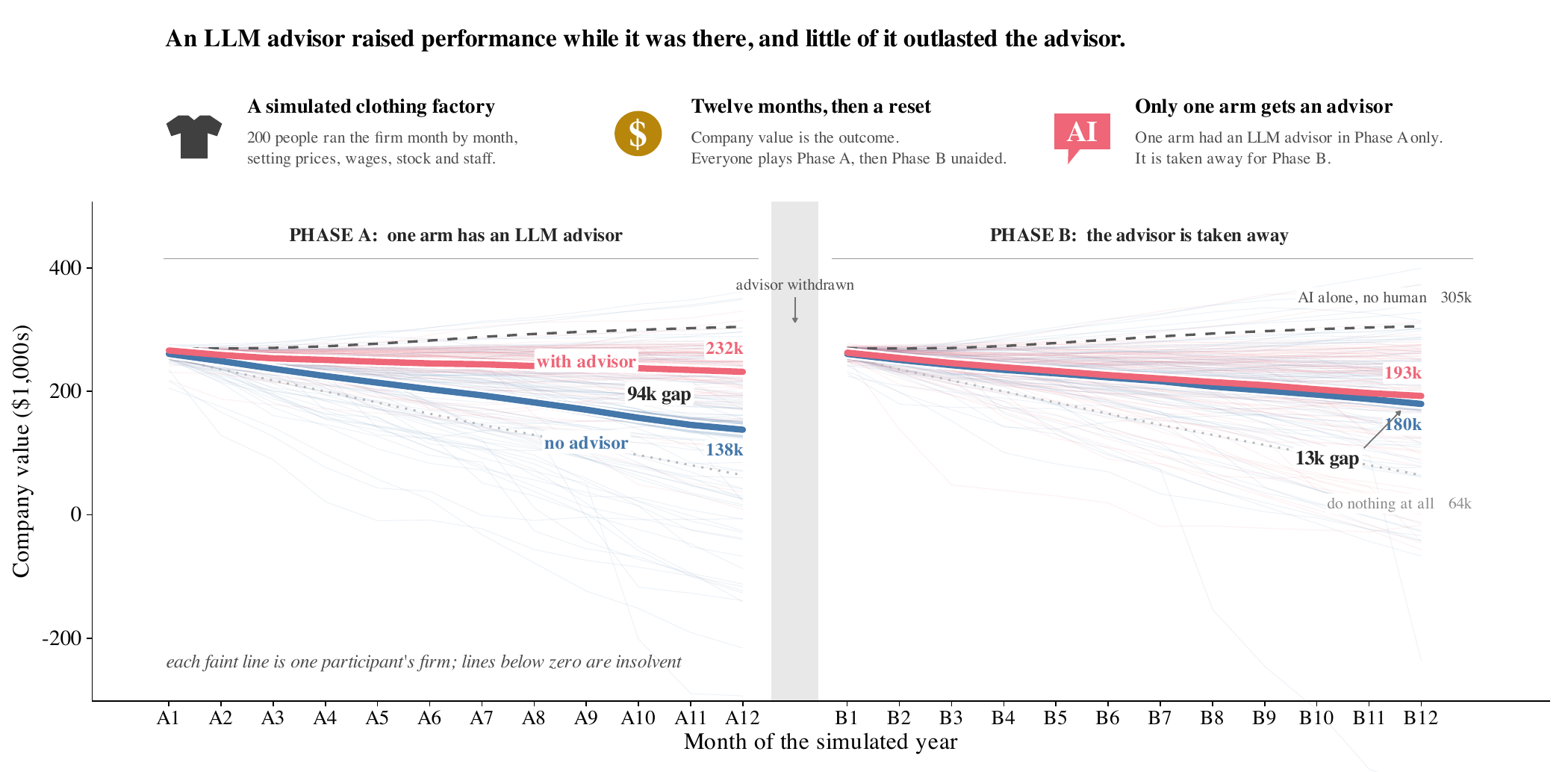}
\caption{\textbf{An LLM advisor raised performance while it was available, and left behind little more than a second unaided run would have.} $200$ participants ran a simulated clothing factory for twelve months with the advisor present for one group (Phase~A), then twelve further months with no advisor for anyone (Phase~B). Faint lines are individual firms; heavy lines indicate the mean with $\pm 1$ standard error of the mean (SEM) ribbons; the dashed line is the mean of $50$ autonomous AI-alone runs, and the dotted line is the do-nothing baseline. The $94$k end-of-phase gap in Phase~A falls to $13$k once the advisor is gone.}
\Description{A wide two-part chart of company value over twenty-four months. A header strip shows three icons: a shirt for the clothing factory, a coin for company value, and a speech bubble marked AI for the advisor. In Phase A, the \condai{} mean stays high while the \condnoai{} mean declines, ending 94 thousand apart. After the advisor is withdrawn at the phase boundary, both means fall together and end 13 thousand apart.}
\label{fig:trajectories}
\end{teaserfigure}

\maketitle
\raggedbottom

\section{Introduction}

A foundational goal of human-computer interaction (HCI) is to engineer technologies that extend human capacity for increasingly complex problems~\cite{engelbart1968research,engelbart1962augmenting,schmidt2020,andy1998extended,clark2010supersizing,clark2025extending}. Artificial intelligence (AI), particularly large language models (LLMs), pushes this vision further: LLMs span unusually broad tasks, are used interactively~\cite{openai2022chatgpt,draxler2023gender}, and are already deployed against real societal problems, though whether that breadth reflects genuine generalization is contested~\cite{gignac2024defining,yiu2024transmission}. They provide observable benefits at scale~\cite{bastani2024generative,dell2023navigating}, though those benefits do not always reach the user's own judgment~\cite{FERNANDES2026108779}.

What makes a problem complex and how people tackle it have been studied in cognitive psychology for four decades~\cite{dorner2017complex,funke2001dynamic}, yet this work is largely absent from the human--AI interaction (HAI) literature. Complex problems are dynamic, opaque, multi-goal systems, studied through computer-simulated microworlds preserving ecological dynamics under laboratory control~\cite{dorner2017complex,funke2001dynamic,danner2011measuring}, unlike the abstract cognitive items (e.g., the Law School Admission Test) used in recent human--AI interaction work~\cite{FERNANDES2026108779}.

Research in complex problem solving (CPS) shows that success rests on the solver's capacity for systems thinking: building a usable mental model of the system and using it to test hypotheses before intervening. That capacity for effective CPS is tied to fluid reasoning but distinct from intelligence~\cite{greiff2015prediction,kretzschmar2021working}. Without such a model, solvers fall into maladaptive patterns such as ``flight into action'', acting rather than modeling~\cite{dorner2017complex,dorner2011logik,dorner1980difficulties}.
Yet HAI research shows three shortcomings in engaging complexity. First, AI-augmentation studies use bounded, checkable tasks (essays~\cite{noy2023experimental}, math coursework~\cite{bastani2024generative}, programming~\cite{Qian2024}, multitasking~\cite{Lingler24}), rather than sustained control of a dynamic system. Second, AI benchmarks evaluate models in isolation on standardized items~\cite{srivastava2023beyond,pellert2023ai,bringsjord2011psychometric,klein2024performance,wang2024mmlu}. Third, the finding that brief AI assistance degrades unaided performance~\cite{liu2026ai} rests on short, static tasks. These strands leave open how assistance affects structural knowledge and later unaided control in a dynamic system.

Assistance can improve joint performance without improving the user's own understanding. Whether it also develops knowledge and unaided control is a separate empirical question, not a requirement that every LLM should teach. We distinguish gains in the user's understanding from gains that depend on assistance during use. Withdrawal tests what users can subsequently do alone, which matters wherever support may become unavailable.

We therefore bring an established CPS microworld, the Tailorshop~\cite{funke2001dynamic,danner2011measuring,brand2024predicting}, to HAI. Participants manage a simulated shirt-manufacturing business for 12 simulated months in Study~1 and 36 in Study~2, adjusting twelve interacting controls each month. We compare humans alone, humans with LLM assistance, and an autonomous LLM on identical tasks. Beyond outcome performance, we measure \emph{systems thinking}, the participant's grasp of the system, using an engine-scored knowledge test administered after the aided phase (H1). Study~1 asked participants to predict how company value responds to specific interventions. Study~2 replaced that format with a $41$-item structural-knowledge battery. Study~1 then withdraws the advisor and tests whether participants can maintain unaided control over the system for an additional 12 months using a reset firm (H2).

We report two preregistered studies. \emph{Study~1} tests H1, the user's model of the system, and H2, unaided control after withdrawal. Its knowledge instrument proved too hard. Both groups scored at chance, so H1 rests on a test that could not detect knowledge where it existed. \emph{Study~2} re-tests H1 alone with an engine-keyed structural-knowledge battery, over an aided phase three times as long and without withdrawal. First, after withdrawal, the advisor's large performance advantage vanished against the do-nothing reference, but a small advantage survived against the harder reference of repeating one's own prior decisions ($g = 0.32$). Second, confidence, experienced effort, and strategy coherence tracked the advisor's presence, not the participant's skill. Assistance thus reproduced the outward signs of effective CPS, while nearly all performance gains vanished upon withdrawal. We term this pattern \emph{capability lending}. Third, knowledge differences were small on the accuracy scale. Study~2's registered contrast put \condai{} slightly ahead ($\theta = +0.42$, $95\%$ credible interval (CrI) $[+0.11, +0.73]$), but that margin largely tracks differential solvency. Unaided participants went bankrupt far more often over Study~2's 36 months. Because solvency is itself an outcome, stratification cannot distinguish learning enabled by survival from selection. Participants who altered at least one recommended control in a larger share of consultations performed better after withdrawal in Study~1 and scored higher on structural knowledge in Study~2.

This work makes four contributions. (i)~We test HAI in a CPS context. (ii)~We measure the user's mental model alongside task performance. (iii)~We characterize from log data when people consult an LLM advisor, how closely they follow its recommendations, which decisions they keep for themselves, and how overriding related to performance. (iv)~We separate what users \emph{know} from what they \emph{believe} they know. Assistance raises the sense of understanding without a corresponding difference on the accuracy scale. The gap is invisible even where there is a right answer. Asked what their own firm was worth, $64\%$ of advised participants placed their unaided outcome at or above what they had achieved \emph{with} help, though it had in fact fallen by $38$k.
Study materials, the minimal anonymized analysis data, and the reproduction notebook, script, and figure code will be released upon acceptance at \url{https://github.com/RobinWelsch/Human-LLM-Dynamic-bench}. Preregistrations are available for \href{https://aspredicted.org/fv7eh6.pdf}{Study~1} and \href{https://aspredicted.org/hf3qz8.pdf}{Study~2}.

\section{Related Work}

\subsection{Complex Problem Solving and Systems Thinking}
Cognitive psychology treats dynamic, ill-structured environments as instances of \emph{complex problem solving}. Five attributes distinguish them from well-defined tasks: many variables (\emph{complexity}), dense causal coupling (\emph{connectivity}), autonomous state evolution (\emph{dynamics}), hidden structure requiring inference (\emph{intransparency}), and conflicting goals (\emph{polytely}). Time is usually short, and interventions are seldom reversible~\cite{dorner2017complex}. Complexity resides in the solver--environment interaction, with no single correct answer~\cite{mirel2004interaction}. To control these properties, researchers built \emph{microworlds} from the 1970s onward: constrained enough for scoring, yet rich enough to demand systems thinking~\cite{dorner2014complex,broadbent1978human}. At one extreme sits Dörner's \emph{Lohhausen} \cite{dorner1980difficulties}, a town of roughly 2{,}000 coupled variables. At the other, scorable scenarios such as the \emph{Tailorshop} \cite{danner2011measuring} and fire-fighting systems~\cite{schoppek2015complex,greiff2015prediction,funke2001dynamic,fischer2017history,brehmer1992dynamic}. A later generation of \emph{minimal complex systems} exchanged ecological richness for psychometric tractability~\cite{greiff2015prediction}. We use the Tailorshop, a microworld that sits between these two extremes.%
\footnote{MicroDYN, the predominant minimal-complex-systems instrument, meets only a subset of the complexity criteria (three inputs and outputs, no side-effects, no conflicting goals)~\cite{schoppek2015complex}.}

CPS performance has two phases with partly distinct abilities: knowledge acquisition, probing the system's causal structure, and knowledge application, exploiting it to reach goal states~\cite{funke2001dynamic,greiff2015prediction,wustenberg2012complex}. The mental model mediates them: structural knowledge strongly predicts later control performance~\cite{funke2014analysis,schoppek2015complex,busswolder2015effect}, and transfer depends on strategy comprehensiveness~\cite{schoppek2017common}. A meta-analysis finds only a moderate correlation with general intelligence ($g$), so CPS is not merely its proxy~\cite{stadler2015complex,kretzschmar2021working}. Mental model quality, the main within-task predictor of success, depends on disciplined acquisition strategies, especially varying one factor at a time (VOTAT). Confounded probing leaves the system's structure unidentified~\cite{vollmeyer1996impact,greiff2012dynamic,lotz2022g}. Without an adequate model, solvers exhibit Dörner's ``flight into action'': intervening too early, repeating successful actions (methodism), ignoring side-effects and long-term consequences, and over-steering under delayed feedback~\cite{dorner1980difficulties,brehmer1992dynamic}. These failures may intensify once the assistant is withdrawn.

CPS research converges on \emph{systems thinking}---understanding how variables interact, accumulate, and feed back over time. Performance is poor by default: even highly educated participants reason about feedback systems in linear, open-loop terms and misjudge simple accumulation~\cite{funke2001dynamic,cronin2009don}. Human factors has long studied this~\cite{sheridan1992telerobotics,brehmer1992dynamic}: automation can \emph{deskill} operators, who cannot resume control at its limits (the \emph{out-of-the-loop} problem~\cite{endsley1999level,bainbridge1983ironies}), yet well-matched support can leave them more capable. An LLM changes the terms: classic automation operated silently within a single context, whereas a foundation model operates across domains via language, explaining mechanisms and potentially teaching. \textbf{This makes the solver's mental model the quantity to measure: participants \emph{anticipate} the system's response to interventions, scored against ground truth.}

\subsection{Augmentation, Offloading, and What the Human Retains}
HCI's augmentation vision~\cite{engelbart1962augmenting,engelbart1968research,schmidt2017augmenting} raises the question of whether assistance leaves the user more capable afterwards, not merely better served during. Studies report sizable productivity gains from LLM assistance on well-structured tasks: writing, tutored mathematics, and programming~\cite{noy2023experimental,bastani2024generative,peng2023impact}. These gains are neither uniform nor self-evidently safe: adding an assistant left software-engineering exam scores unchanged overall and helped only novices on open-ended items~\cite{Qian2024}, and the comparable multitasking study used a reinforcement-learning scheduler, not an LLM~\cite{Lingler24}. Three features distinguish our CPS setting from these tasks: an extended horizon, conflicting goals, and autonomous dynamics with feedback to model.

AI assistance can also \emph{degrade} learning: in the math study, students who practiced with GPT scored 17\% \emph{worse} than controls once it was removed~\cite{bastani2024generative}. A ``GPT Tutor'' group eliminated this harm (null vs.\ control), suggesting that tool design moderates substitution. Cognitive forcing functions that reduce over-reliance are resisted by users who would benefit most~\cite{buccinca2021trust}. Broader work covers automation-induced deskilling~\cite{rafner2022deskilling} and professionals' sense of contribution~\cite{Kobiella2024}. A meta-analysis of over one hundred human--AI experiments finds the team underperforms its better member on average, except on creation tasks or when the human is stronger~\cite{vaccaro2024combinations}. Whether the assisted human can continue unaided on a task that is dynamic, intransparent, and polytelic remains an open question. \textbf{What users retain after AI-supported complex problem solving remains an open question, distinct from what the pair produces.\footnote{For AI research on CPS without human--AI interaction, see~\cite{zheng2025knowledge}}}

This gap has a well-established mechanism: when an external store holds part of a task, people offload rather than encode it, so its removal reveals shallower memory~\cite{risko2016cognitive,kelly2019isolation,kelly2025does,kelly2022study}. With AI, this appears as over-reliance on unchecked output~\cite{vasconcelos2023explanations}, automation bias and complacency~\cite{parasuraman2010complacency}, and confidence detached from accuracy~\cite{colombatto2023illusions,colombatto2025metacognition,tankelevitch2023metacognitive,FERNANDES2026108779}.
The converse is equally plausible: an advisor showing its reasoning may serve as a worked example, building rather than replacing the user's mental model. Both scenarios feel similar during assistance (greater control, less effort) but differ in residual knowledge and capability once the advisor is withdrawn.
\textbf{Outcome metrics in human--AI interaction cannot distinguish between these scenarios, and they are almost always collected with the assistant still in place. The critical assessment is what the user knows and can perform unaided after AI assistance is withdrawn.}

\section{Study~1}\label{sec:s1-method}
\begin{figure}[!htbp]
\centering
\resizebox{\linewidth}{!}{%
\begin{tikzpicture}[
 font=\sffamily\small,
 x=1cm, y=-0.78cm,
 shared/.style ={draw, fill=gray!10, align=center, minimum width=10.8cm, minimum height=7.5mm, inner sep=4pt},
 solo/.style ={draw, fill=white, align=center, minimum width=5.3cm, minimum height=7.5mm, inner sep=4pt},
 withai/.style ={draw, fill=cpsai!30, align=center, minimum width=5.3cm, minimum height=7.5mm, inner sep=4pt},
 withdraw/.style ={draw, fill=orange!30, align=center, minimum width=5.3cm, minimum height=7.5mm, inner sep=4pt},
 measure/.style ={draw, fill=yellow!40, align=center, minimum width=10.8cm, minimum height=7.5mm, inner sep=4pt, font=\bfseries\small},
 notice/.style ={draw, dotted, thick, fill=orange!12, align=center, minimum width=5.3cm, minimum height=6mm, inner sep=3pt, font=\itshape\footnotesize},
 comp/.style ={draw, fill=green!15, dashed, align=center, minimum width=15.6cm, minimum height=8.5mm, inner sep=4pt},
 timing/.style ={font=\footnotesize\itshape, gray!50!black, align=center, minimum width=1.8cm, anchor=east},
 rmeas/.style ={font=\footnotesize, align=left, anchor=west, text width=4.6cm},
 arr/.style ={->, >=Latex, semithick}
]

\node[font=\bfseries] at (-2.65, -0.7) {\condnoai};
\node[font=\bfseries] at ( 2.65, -0.7) {\condai};
\node[font=\bfseries\footnotesize, gray!50!black, anchor=east] at (-5.6, -0.7) {time};
\node[font=\bfseries\footnotesize, gray!40!black, anchor=west] at ( 5.6, -0.7) {measures collected};

\node[timing] at (-5.6, 0) {$\sim$3\,min};
\node[shared] (consent) at (0, 0) {Consent + demographics (age, gender, education, AI experience)};

\node[timing] at (-5.6, 1) {$\sim$2\,min};
\node[shared] (att) at (0, 1) {Attention check (instructed-response + comprehension items)};

\node[timing] at (-5.6, 2) {$\sim$5\,min};
\node[shared] (ncs) at (0, 2) {Need for Cognition Scale (NCS-6) and Cognitive Reflection Test (CRT)};
\node[rmeas] at (5.6, 2) {NCS-6 score; CRT score (descriptive)};

\node[timing] at (-5.6, 3.2) {$\sim$2\,min};
\node[shared] (cover) at (0, 3.2) {Tailorshop cover story (no mechanism hints)};

\node[timing] at (-5.6, 4.2) {$\sim$3\,min};
\node[shared] (tut) at (0, 4.2) {Tailorshop interface and controls walkthrough (no mechanism hints; prediction task announced)};

\node[timing] at (-5.6, 5.2) {$\sim$6\,min};
\node[shared] (expl) at (0, 5.2) {Tailorshop \textbf{exploration}: 6 months, unaided, unscored (both groups; free experimentation, unprimed by any elicitation)};

\node[timing] at (-5.6, 7.8) {$\sim$25\,min};
\node[solo] (paN) at (-2.65, 7.8) {\textbf{Phase~A} -- months 1--12\\ no advisor};
\node[withai] (paA) at ( 2.65, 7.8) {\textbf{Phase~A} -- months 1--12\\ advisor in chat panel};
\node[rmeas] at (5.6, 7.8) {Per round: decision confidence, mental effort, understanding, control, AI reliance (\condai{} only); company value logged monthly.};

\node[timing] at (-5.6, 9.3) {$\sim$8\,min};
\node[measure] (post) at (0, 9.3) {Prediction task\hspace{0.6em}{\normalfont(24 engine-scored items; once, after Phase~A)}};
\node[rmeas] at (5.6, 9.55) {\textbf{Prediction accuracy} (share of 24 items correct): the mental-model measure};

\node[timing] at (-5.6, 10.5) {$\sim$3\,min};
\node[withai] (trustb) at ( 2.65, 10.5) {\condai{} only: Trust in Automated Systems (TIAS, 12 items) + 2 AI-reliance items};

\node[notice] (notice) at ( 2.65, 11.7) {``The assistant is unavailable for the remaining 12 months.''};

\node[timing] at (-5.6, 12.9) {$\sim$25\,min};
\node[solo] (pbN) at (-2.65, 12.9) {\textbf{Phase~B} -- 12 mo.\ (reset)\\ no advisor};
\node[withdraw] (pbA) at ( 2.65, 12.9) {\textbf{Phase~B} -- 12 mo.\ (reset)\\ \textbf{advisor withdrawn}};
\node[rmeas] at (5.6, 12.9) {Per round: Phase-A battery minus the AI-reliance item; company value logged monthly.};

\node[timing] at (-5.6, 14.1) {$\sim$4\,min};
\node[measure] (ret) at (0, 14.1) {Manipulation check (both groups)};
\node[rmeas] at (5.6, 14.45) {Manipulation check: was an AI advisor available? (both groups); usage frequency if yes};

\node[timing] at (-5.6, 15.3) {$\sim$5\,min};
\node[shared] (surv) at (0, 15.3) {Strategy survey: open-ended responses + per-variable relevance ratings};

\node[timing] at (-5.6, 16.3) {$\sim$2\,min};
\node[shared] (debrief) at (0, 16.3) {Debrief (study purpose disclosed) + suspicion probe + Prolific completion redirect};

\node[font=\bfseries\footnotesize, anchor=east] at (-3.7, 17.3) {Total session $\approx$ 90 min};

\node[comp] (comp) at (0, 18.7) {%
 \textbf{AI-alone benchmark} -- runs in parallel; no participants, no chat.\\
 Same LLM, same system prompt as the advisor. Acts on the simulation via the same structured action interface available to humans.\\
 50 runs on the single default market seed shared by all human sessions. Used as the descriptive AI-alone reference.%
};

\foreach \a/\b in {consent/att, att/ncs, ncs/cover, cover/tut, tut/expl, surv/debrief}
 \draw[arr] (\a) -- (\b);

\draw[arr] (expl.south -| paN.north) -- (paN.north);
\draw[arr] (expl.south -| paA.north) -- (paA.north);

\draw[arr] (paN.south) -- (post.north -| paN.south);
\draw[arr] (paA.south) -- (post.north -| paA.south);

\draw[arr] (post.south -| trustb.north) -- (trustb);
\draw[arr] (trustb) -- (notice);
\draw[arr] (notice) -- (pbA);
\draw[arr] (post.south -| pbN.north) -- (pbN.north);

\draw[arr] (pbN.south) -- (ret.north -| pbN.south);
\draw[arr] (pbA.south) -- (ret.north -| pbA.south);
\draw[arr] (ret) -- (surv);

\end{tikzpicture}%
}
\caption{%
\textbf{Both groups share every stage; the conditions diverge only in the advisor's presence during Phase~A and its withdrawal for the unaided Phase~B.} Study flow with timing, condition contrast, per-stage measures, and the autonomous-AI baseline.
\textbf{Lanes.} Grey boxes span both groups; white = unaided play, coral = LLM advisor available, orange = withdrawal phase (advisor withdrawn). The yellow box is the prediction task (24 engine-scored items), administered once, immediately after Phase~A.
The dotted box on the \condai{} lane between Phase~A and Phase~B is the single withdrawal-notice screen shown to participants at the start of Phase~B.
\textbf{Time column} (left) gives approximate stage duration; the total session runs $\sim$90 minutes.
\textbf{Measures column} (right) lists the data collected at each stage: descriptive individual-difference scales, the post-Phase-A prediction task, per-round self-report and company value across the 24-month test, the \condai{}-only trust-and-reliance battery (administered after the aided phase, before withdrawal), and the post-task strategy survey and debrief.
The \textbf{AI-alone benchmark} (dashed green) uses the same LLM and prompt as the advisor, acting on the simulation via the same structured action interface, run on the same fixed default market seed as every human session.%
}
\label{fig:flow}
\Description{A two-lane study-flow timeline. Both groups share an exploration phase, the post-Phase-A prediction task, and individual-difference scales; the \condai{} lane additionally completes a trust-and-reliance battery after the aided phase, and both complete a post-task strategy survey and debrief. The two lanes diverge in Phase~A (the \condai{} lane has the LLM advisor available, the \condnoai{} lane plays unaided) and both play Phase~B unaided after a withdrawal-notice screen on the \condai{} lane. A separate dashed lane shows the AI-alone benchmark run on the same seeds.}
\end{figure}
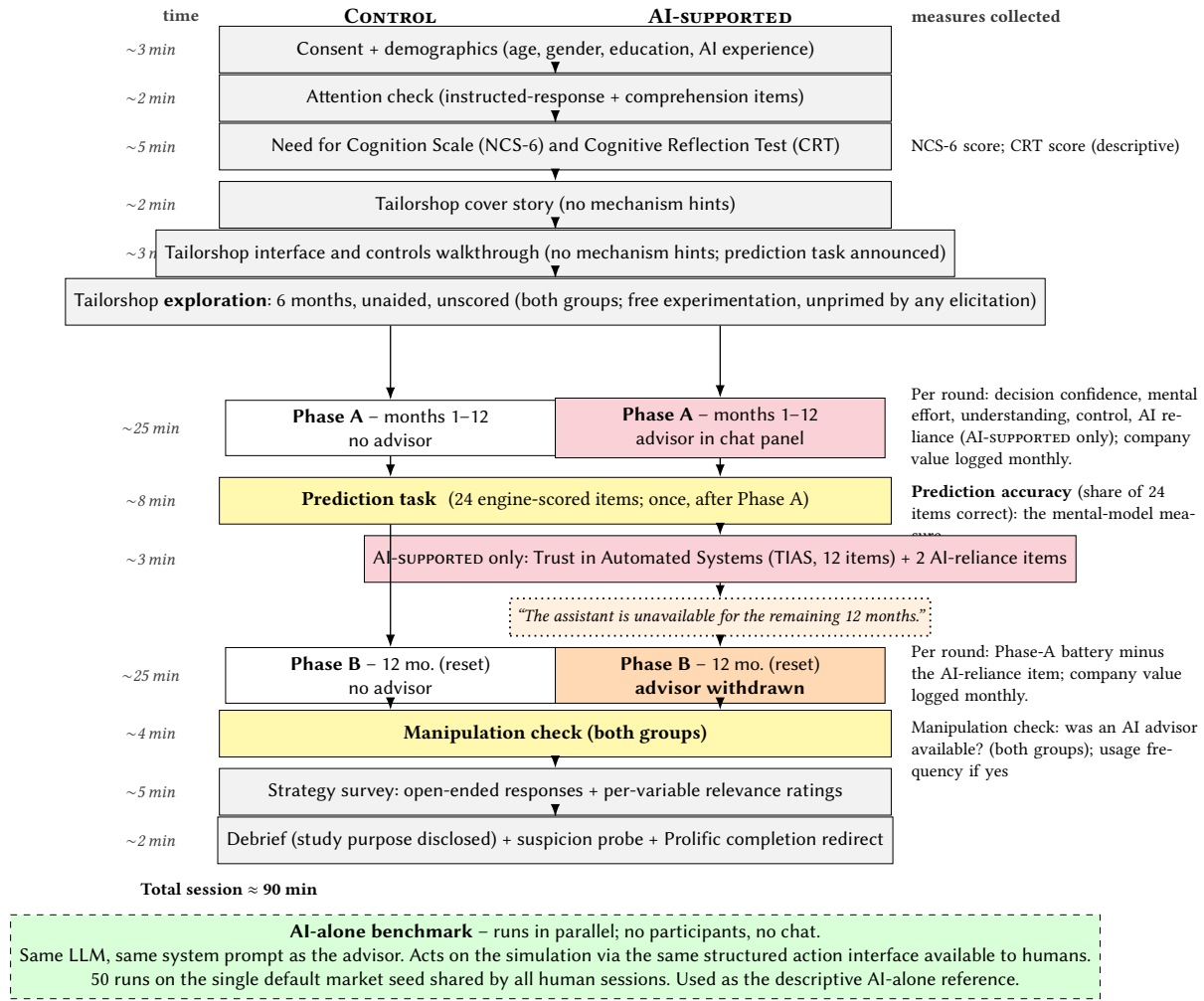

Participants in \condai{} used an LLM advisor during a 12-month Tailorshop run (Phase~A). Both groups then completed 12 unaided months after the firm was reset and the advisor removed (Phase~B). Two confirmatory, two-sided hypotheses are tested. \textbf{H1 (between participants):} prior advisor availability changes the accuracy of the user's model of the system, measured by a prediction task given immediately after Phase~A, with \condai{} below \condnoai{} as the substitution direction. \textbf{H2 (withdrawal):} the between-group gap in decision quality, measured by the state-conditional beat rate (\autoref{sec:measures}), changes from aided Phase~A to unaided Phase~B. The rate scores each month against a do-nothing counterfactual from the participant's own state, anchoring it to the engine's dynamics by construction.
These outcomes are compared against the preregistered verdict table (\autoref{tab:verdict}). Five exploratory analyses follow: \textbf{E1} in-task benefit, \textbf{E2} reliance dose (uninformative: uptake was self-selected and near-saturated, so the dose--response probe has almost no variance to work with; the one usable contrast, self-reported reliance against unaided Phase-B rate, is in \autoref{tab:stratification}), \textbf{E3} unaided strategy, \textbf{E4} advisor-contingent confidence, and \textbf{E5} self-assessed outcome. Each confirmatory hypothesis has a preregistered verdict rule, so no multiplicity correction is applied. Exploratory $p$-values are uncorrected. Interpretation emphasizes effect sizes and intervals.
\begin{table}[t]
\caption{\textbf{The verdict table diagnoses augmentation and practical equivalence on equal footing with substitution.} Pre-registered joint verdict over the two confirmatory dependent variables (DVs). Each DV is judged \emph{worse} / \emph{equivalent} / \emph{better} for \condai{}: its confirmatory mixed model gives the \emph{direction} (for H1 the between-group prediction-accuracy contrast, for H2 the phase$\times$condition withdrawal interaction, where \emph{worse for ai} means the state-conditional gap narrows on withdrawal, not that unaided performance falls below controls (the registered plan reports the Phase-B condition effect alongside it)) and the Bayesian counterpart's $95\%$ highest-density interval (HDI) lying within the $\pm 0.40$ region of practical equivalence (ROPE) renders the verdict as \emph{equivalent}; a DV that is neither, is \emph{inconclusive}, making the overall verdict inconclusive. Bold cells are the three theory-anchored outcomes. For Study~1 the H1 axis is \emph{void} rather than \emph{equivalent}: both groups answered that battery at chance, so the equivalence rule cannot be applied to it and the joint verdict is inconclusive on the model axis which motivated Study 2.}
\Description{A three-by-three verdict table crossing the H1 mental-model outcome (worse, equivalent, better for \condai{}) with the H2 withdrawal outcome, assigning each cell a reading such as substitution, augmentation, no measurable cost, or mixed.}
\label{tab:verdict}
\footnotesize
\setlength{\tabcolsep}{4pt}
\begin{tabular}{@{}p{0.18\linewidth}p{0.23\linewidth}p{0.23\linewidth}p{0.23\linewidth}@{}}
\toprule
 & \multicolumn{3}{c}{\textbf{Withdrawal of the state-conditional rate (phase$\times$condition)}}\\
\cmidrule(l){2-4}
\textbf{Prediction accuracy (mental model)} & worse for ai & equivalent & better for ai \\
\midrule
worse for ai & \textbf{substitution} & substitution (model) & mixed \\
equivalent & performance cost only & \textbf{no measurable cost} & augmentation (transfer) \\
better for ai & mixed & augmentation (model) & \textbf{augmentation} \\
\bottomrule
\end{tabular}
\end{table}
The design dissolves the human--AI composite and measures what remains once the human faces the task without the AI: a 2 (Condition: \condnoai{} vs.\ \condai{}; between-subjects) $\times$ 2 (Phase: A vs.\ B; within-subjects) mixed design (\autoref{fig:flow}). \condnoai{} participants never had an advisor. \condai{} participants had one in a chat panel throughout Phase~A, and the advisor was withdrawn for Phase~B. 

We preregistered the study on AsPredicted (\#301{,}976, \url{https://aspredicted.org/fv7eh6.pdf}).\footnote{Timestamped 2026-07-15 05:47~PT, $7.1$~hours before the first session; no screened session predates it.} Upon acceptance, we will release the minimal anonymized analysis data, study and analysis code, and model configuration at \url{https://github.com/RobinWelsch/Human-LLM-Dynamic-bench}. Participant identifiers, free-text responses, transcripts, and raw telemetry will not be released. The study required no ethical review under the \href{https://tenk.fi/en/advice-and-materials/guidelines-ethical-review-human-sciences}{Finnish National Board on Research Integrity (TENK) guidelines}.

\subsection{Participants and power}\label{sec:power}

The a priori target of $N = 200$ ($100$ per group) was determined by Monte Carlo simulation of the two confirmatory models, with nuisance parameters fixed in advance. At $N = 200$ the simulation gives H1 $.84$ power for an $8$-point accuracy gap and $.98$ for a $10$-point gap. The H2 withdrawal interaction has $.91$ power for a conservatively reduced log-odds effect of $-0.45$ (parameters in \autoref{app:power}; cell-by-cell tables and the ROPE decision-rate simulation in the materials release).
Recruitment continued until 200 sessions included both primary outcomes and passed the applied exclusions (deviations D1, D4, and D5; \autoref{app:deviations}). Of 232 Prolific completers, 222 provided a complete prediction battery and 231 completed Phase~B. Ten completed only Phase~B and one only the battery. The analytic sample comprised 100 participants per group, aged 18--67, with mean ages of 33.5 (\condnoai{}) and 33.9 (\condai{}). It included 110 men, 89 women, and one non-binary participant. A bachelor's degree was modal in both groups. The applied screening rules excluded $17$ \condnoai{} and $4$ \condai{} participants overall (Fisher's exact test, $p = .010$, odds ratio $4.18$). 

Compensation was about GBP~10 per hour against a $\sim$90-minute target. The seven preregistered exclusion criteria screen on attention and comprehension items, the manipulation check, the registered session cap, technical failure or non-engagement, a hidden attention honeypot, an implausible company-value forecast, and a median per-decision latency below $15$\,s. Each is operationalized in \autoref{app:method-details}.

\subsection{Apparatus and Software}\label{sec:apparatus}

\begin{figure}[!htbp]
\centering
\includegraphics[width=\textwidth]{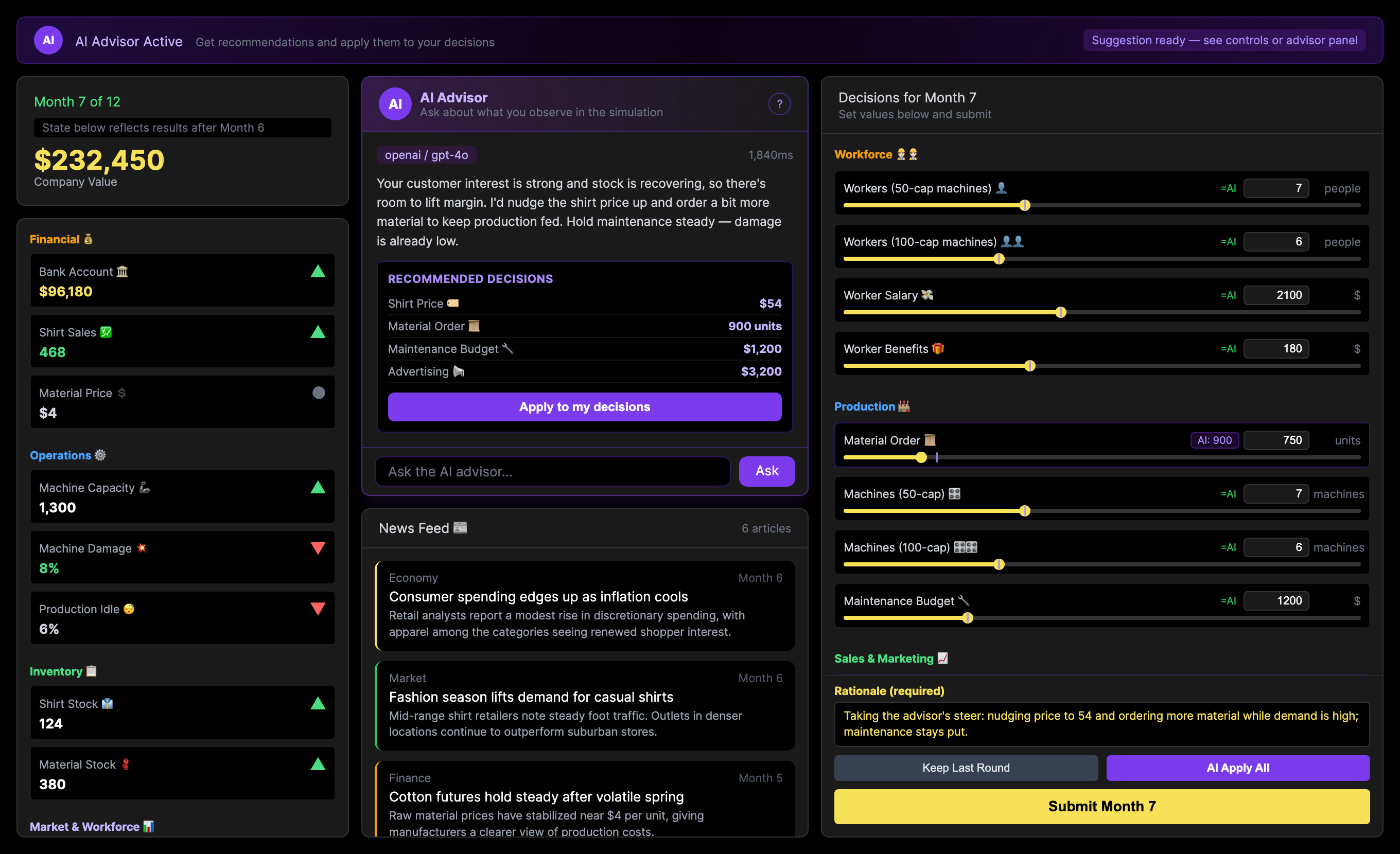}
\caption{\textbf{Both conditions share one Tailorshop interface; the advisor is the only thing
added.} The monthly decision dashboard as \condai{} participants saw it in Phase~A: the left
column reports the firm's state (company value in the brand accent, plus the coupled economic
variables with per-month trend arrows), the right column holds the twelve decision controls
submitted each month, and the center a scrolling news feed. Three elements are added over the
identical game: an ``AI Advisor Active'' banner, a queryable advisor panel that rendered the advisor's full reply (analysis, options, recommendation and risks as text and option cards; the suggested decisions as per-control chips with an \emph{Apply} action; replies ran to a median of $343$ displayed words), and per-control
suggested-value chips, while the underlying simulation, seed, and state are unchanged.
\condnoai{} participants saw this same dashboard with those three elements absent. Screens are faithful reconstructions of the deployed frontend with
illustrative values, not participant data.}
\Description{A screenshot of the Tailorshop management dashboard in the \condai{} condition,
showing a left panel of company-state variables, a right panel of twelve monthly decision
controls, and a central news feed, plus a purple ``AI Advisor Active'' banner, an advisor
recommendation panel with an Apply button, and suggested-value chips on the decision controls.}
\label{fig:ui}
\end{figure}

The study ran in participants' desktop browsers (mobile access was blocked). \autoref{fig:ui} shows the \condai{} dashboard. The \condnoai{} one is identical with the advisor elements removed and is reconstructed in the materials release.
The advisor informed every decision but executed none. It saw only what the participant saw, and its recommendations reached the simulation only upon submission (\autoref{fig:advisor}).

\subsection{Materials}

\subsubsection{Tailorshop microworld.}\label{sec:tailorshop}
The Tailorshop is a managerial microworld of a small shirt-manufacturing business. Twelve monthly decision controls govern a 12-variable state vector under nonlinear, coupled dynamics. Both groups used identical seeds, and without skilled intervention the firm steadily loses value. The couplings that make a model necessary are delayed or indirect: demand follows customer interest and retail price nonlinearly, interest itself responds to advertising and outlets, and machine capacity decays unless maintained. The state vector, the twelve controls and the full parameterization are in \autoref{app:method-details}. Play ran in three phases: a 6-month unscored unaided exploration phase, then Phases~A and~B.

\subsubsection{Prediction task.}
An engine-scored \emph{prediction task} immediately after Phase~A measured the mental model. It included 24 items, each pairing a company state with a candidate decision and asking whether company value would be \emph{higher} or \emph{lower} three months later than with no change (forced choice, five-point confidence rating). Items cover the marketing, production, and human-resources subsystems, several dynamically counter-intuitive, and split into intuition-aligned and misaligned halves. The engine keys each by replaying the queried decision against the do-nothing default from the same state.

\subsubsection{LLM advisor.}

\autoref{fig:advisor} summarizes the consultation process. The advisor was OpenAI \texttt{gpt-5.5} (one pinned snapshot, shared with the autonomous benchmark),
queried with a fixed system prompt (\autoref{app:prompts}) that instructed it to act as an advisor and had no access to the internal variables of the Tailorshop engine. It responded only when consulted, at reasoning effort \texttt{none} with a $1{,}024$-token completion budget and fixed default sampling. The panel showed only the latest exchange, while the model retained the full Phase-A transcript.

\subsubsection{AI-alone benchmark.}
The benchmark used the same model and task description but acted through a structured action interface (prompts and turn template in the data release): $r = 50$ sessions on the same fixed market, both phases starting from the same reset. Decoding matched the advisor's: reasoning effort \texttt{none} with a
$1{,}024$-token completion budget. The benchmark \emph{increased} the firm's value (rising-months rate $0.91$, end value $305$k), unlike the typically declining human trajectories reported across Tailorshop samples~\cite{danner2011measuring,danner2011beyond,brand2024predicting}.

\subsection{Procedure}
After consent, participants completed demographics, a four-item attention and comprehension check with a hidden honeypot, the six-item Need for Cognition Scale (NCS-6)~\cite{lins2020very}, a seven-item Cognitive Reflection Test (CRT)~\cite{sirota2018effect}, a cover story, and a prediction-task walkthrough. Conditions were identical until this point (\autoref{fig:flow}). Each month, participants provided self-reports and a required free-text rationale. In \condai{}, the advisor panel was cleared each month, and users could apply recommendations with one click (\emph{Apply all}). Median play lasted $36$ minutes within an approximately $90$-minute session.

\subsection{Measures}\label{sec:measures}
H1 and H2 use prediction-task scores and logged engine states.

\begin{figure}[!htbp]
\centering
\includegraphics[width=\textwidth]{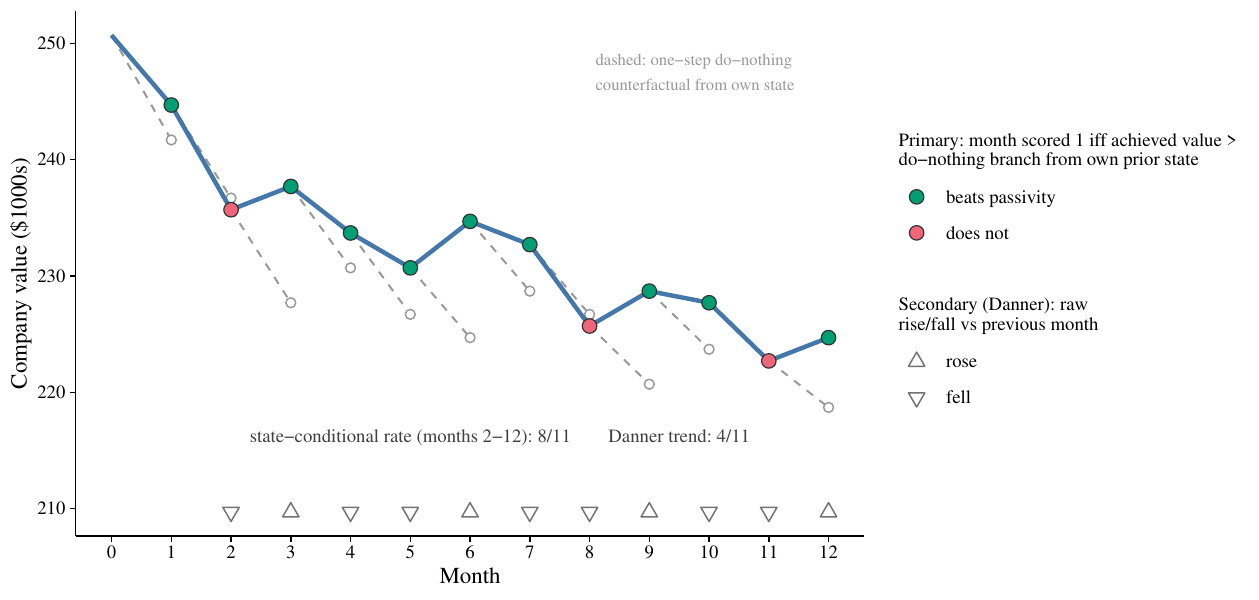}
\caption{\textbf{Scoring each month against a do-nothing counterfactual credits skilled decisions in bad situations that the raw rising-months rate penalises.} How performance is scored (schematic, illustrative data). \emph{Primary (state-conditional):} at every month the participant's recorded state is branched once with the pre-filled do-nothing defaults (dashed segments); the month is scored 1 (filled green) iff the achieved company value exceeds that counterfactual, 0 otherwise (filled red). The H2 rate is the proportion of scored months (2--12). \emph{Secondary (rising-months rate):} the raw rise/fall of company value versus the previous month (triangles). The example shows why the primary is fairer to good decisions in bad states: a steadily declining trajectory scores only 4/11 on the raw rising-months rate but 8/11 against passivity: most of its months are the best of a bad situation.}
\Description{A single declining company-value trajectory over twelve months. From each month's point, a short dashed segment shows where the value would have gone under do-nothing defaults; each achieved point is coloured green when it lies above that counterfactual and red when it does not. A strip below marks each month's raw rise or fall as up or down triangles. The trajectory falls in most months (rising-months rate 4 of 11) yet beats the do-nothing counterfactual in 8 of 11 months.}
\label{fig:danner_schematic}
\end{figure}

\paragraph{Performance (the state-conditional rate).} Performance builds on the rising-months rate of Danner et al.\ \cite{danner2011measuring}: the proportion of months 2--12 in which company value rose over the prior month~\cite{funke1983einige} (\autoref{fig:danner_schematic}). The primary registered outcome is the \emph{state-conditional rate}: recorded decisions are replayed to reconstruct each month's state, the engine is branched once with do-nothing defaults, and the month scores 1 if the achieved value exceeds that counterfactual, a deterministic comparison from the same recorded state, though groups encounter different state distributions. Psychometric properties are in the materials release; \autoref{app:method-details} reports a persistence sensitivity (scored against the participant's own previous-month decisions), the rising-months rate as a secondary comparability measure, and end-of-phase value as descriptive.

\paragraph{Mental model (prediction accuracy).} H1 measures the proportion of the 24 prediction items answered correctly after Phase~A: the ability to anticipate the system's response to an intervention, distinct from controlling it (H2). Internal consistency was low with Kuder--Richardson 20 (KR-20) $= .56$ and $\omega_t = .77$~\cite{revelle2019reliability}. Item-weighting and exclusion sensitivities assess whether problematic items affect the contrast. Metacognitive sensitivity is each participant's mean confidence on correct minus incorrect answers (here months with rising company value); positive values indicate that confidence discriminates response accuracy.

\paragraph{Process measures.} Intervention rate (mean decision controls changed per month) and variable diversity (distinct controls touched across Phase~B, 0--12).

\paragraph{Self-report.} Each round, participants rated decision confidence (0--100 visual analogue scale)~\cite{FERNANDES2026108779}, mental effort (1--9)~\cite{paas1992training}, system understanding (1--7), perceived control (1--7), and AI reliance (1--7; \condai{} during Phase~A only). Post-task trust and strategy measures had no registered hypotheses and are omitted here for brevity (\autoref{app:method-details}); reported reliance quantities are defined below.

\paragraph{Reliance quantities.} Three reliance measures recur and are not interchangeable: the lenient binary \emph{follow-rate} (share of consulted months whose recommendation was applied; median $.99$), the continuous \emph{adherence index} (share of the twelve controls set within $15\%$ of the recommended value, averaged over recommended months; median $.90$), and the \emph{recommendation-level follow proportion} (share of recommended months with adherence $\geq .9$; $86\%$).

\paragraph{Individual differences (descriptive only).} NCS-6 total, CRT total, age, and prior AI-tools experience (1--5) were not used as covariates in the confirmatory tests.

\subsection{Analysis Plan}\label{sec:analysis-plan}

The confirmatory models are mixed-effects logistic regressions: \textbf{H1}, $\text{correct} \sim \text{condition} + (1\,|\,\text{participant}) + (1\,|\,\text{item})$. \textbf{H2}, $\text{beat} \sim \text{phase} \times \text{condition} + (1\,|\,\text{participant})$, over months~2--12 of both phases. Tests are two-sided at $\alpha = .05$, despite directional predictions, using likelihood-ratio tests in R (\texttt{lme4}~\cite{bates2015package}, \texttt{brms}~\cite{burkner2017brms}). Exploratory self-reports use linear mixed models with participant random intercepts and Satterthwaite degrees of freedom (\texttt{lmerTest}~\cite{kuznetsova2017lmertest}). Registered practical equivalence requires a non-significant test and a Bayesian 95\% interval within the $\pm 0.40$ region of practical equivalence (ROPE). This broad threshold does not exclude smaller, potentially relevant effects. Exploratory contrasts report Hedges' $g$ (\condai{} minus \condnoai{}) with 95\% confidence intervals (CIs), using Welch's $t$ or Mann--Whitney $U$ after a Shapiro--Wilk residual check, and Fisher's exact test for binary outcomes.

Analyses use the same 200 participants with both primary outcomes (deviation D4). Missing outcomes are not imputed. Sensitivities assess per-test inclusion and the exclusion departures (D1 and D5). Full Bayesian specifications, WAMBS reporting~\cite{depaoli2017improving}, and deviations appear in \autoref{app:method-details} and \autoref{app:deviations}.

\section{Study~1: Results}\label{sec:prelim}
\subsection{Phase A performance: the pair beat the human and lost to the AI}\label{sec:aided}
\condai{} beat the do-nothing counterfactual in $88.4\%$ of months versus $73.4\%$ for \condnoai{} ($b = +1.09$ log-odds, $z = 5.14$, $p < .001$, odds ratio $2.97$; between-person $g = +0.76$, $95\%$ CI $[+0.47, +1.05]$). The pair ended Phase A at $232$k versus $305$k for the model alone, and $92\%$ of aided participants finished below the worst of its fifty runs. On this task, the human and the AI do not exhibit synergy~\cite{vaccaro2024combinations,bansal_does_2021}: the human added error to a better decision-maker, worsening the outcome. AI recommendations beat passivity in $92\%$ of consulted months and participants followed $86\%$ of them indiscriminately: losing advice was followed as often as the rest, and the $123$ overrides usually cost them ($72\%$ of overrides yielding less value than replaying the recommendation).
The advisor raised the Phase-A mean end company value ($138$k for \condnoai{} versus $232$k for \condai{}; Mann--Whitney $p < .001$, rank-biserial $r = +.49$) and cut outcome variability to about one-third (standard deviation (SD) $= 132$k versus $48$k; $p < .001$). Only two \condai{} participants finished below the do-nothing baseline versus twenty-six in \condnoai{}, and none went bankrupt, whereas sixteen did in \condnoai{} (Fisher's exact test, $p < .001$). The advantage in value retained versus passivity reached $g = +0.94$, $95\%$ CI $[+0.65, +1.23]$, consistent with fewer Phase-A bankruptcies.

\subsection{The user's model (H1).} 
\begin{figure}[!htbp]
\centering
\includegraphics[width=\textwidth]{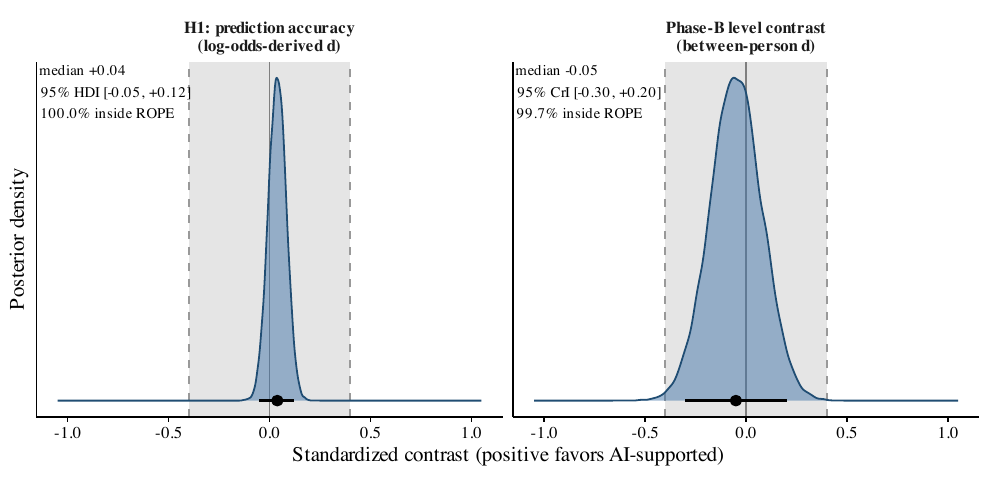}
\caption{\textbf{Study~1: prediction and unaided-level contrasts lie inside the region of practical
equivalence.} Posteriors of the standardized between-group contrast, with the $\pm 0.40$ ROPE
shaded and $95\%$ intervals as black bars (H1 HDI; Phase-B equal-tail CrI). H1 uses a log-odds-derived standardization, Phase~B a probability difference divided by the between-person SD. Positive values favor \condai{}. H1 holds $100\%$ of its posterior mass inside the ROPE and H2's Phase-B level $99.7\%$. That licenses a \emph{practically equivalent} reading of the Phase-B \emph{level}, which is not the withdrawal interaction the verdict table crosses on its H2 axis. On that axis the study is \emph{worse for \condai{}}. It does not on H1: a battery both groups answered at chance cannot register the knowledge the rule presumes, so that axis is void rather than equivalent. Mass inside the band means no
effect as large as $d = 0.40$. It does not exclude smaller ones. Both panels are beta-binomial
models on per-participant counts.}
\Description{Two density plots of standardized effect sizes with a shaded band from minus 0.4 to
plus 0.4. Both densities are centered near zero and sit almost entirely inside the band.}
\label{fig:rope_s1}
\end{figure}

No between-group difference in prediction accuracy was detected: $0.495$ (\condnoai{}) versus $0.512$ (\condai{}; condition effect $+0.08$ log-odds, 95\% CI $[-0.10, 0.27]$, $\chi^2(1) = 0.77$, $p = .38$). The Bayesian contrast is $d = +0.04$ (95\% highest-density interval (HDI) $[-0.05, 0.12]$), entirely within the $\pm 0.40$ ROPE (\autoref{fig:rope_s1}). A two-parameter logistic (2PL) model allowing items to differ in discrimination gives $+0.05$ (95\% CrI $[-0.25, +0.35]$; $99\%$ within the ROPE). Thus, item weighting does not change the result. However, neither group exceeds chance, and both fall below the $.625$ attainable by always answering \emph{lower}. Response-bias and item-level checks indicate that the battery poorly distinguishes knowledge; restricting it to stronger-effect items does not reveal a group difference (\autoref{app:robustness-details}). The result therefore cannot establish equivalent knowledge, and the registered intuition-aligned versus misaligned split is likewise uninformative.

\subsection{Unaided performance (H2).} 
\begin{figure}[!htbp]
\centering
\includegraphics[width=\textwidth]{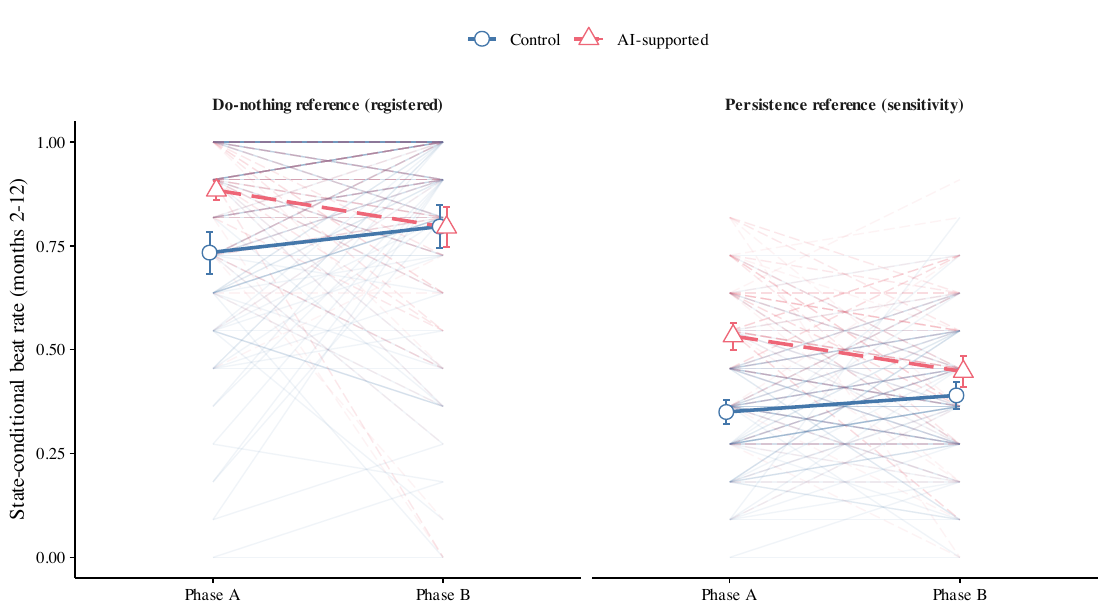}
\caption{\textbf{The advisor's performance advantage is gone by the first unaided month, not eroded gradually ($n=200$; the both-outcomes analytic sample).} State-conditional beat rate (months 2--12) by phase and condition under the registered do-nothing reference (left) and the harder persistence-reference sensitivity (right); points show condition means with 95\% confidence intervals across participants (proportion scale), thin lines individual participants. The Phase-A advantage of \condai{} disappears in the unaided Phase~B on the registered do-nothing reference; on the ceiling-free persistence reference, a small \condai{} advantage survives ($b = +0.24$, $p = .022$, $g = 0.32$).}
\Description{Two interaction panels. The AI-supported mean starts above Control in Phase A. In Phase B the means converge under the do-nothing reference, while a smaller AI-supported advantage remains under the persistence reference.}
\label{fig:h2_interim}
\end{figure}

The confirmatory phase$\times$condition interaction tests the change in the aided advantage at withdrawal (\autoref{fig:h2_interim}): $79.6\%$ \condnoai{} versus $79.5\%$ \condai{} in Phase~B; interaction $-1.23$ log-odds, 95\% CI $[-1.56, -0.89]$, $\chi^2(1) = 51.67$, $p < .001$, odds ratio $0.29$. Its direction persists under both registered alternative scorings, a participant random-slope model, and restrictions to firms remaining above zero or $50$k in Phase~A (\autoref{app:robustness-details}).
Within-group changes capture unaided practice gains in controls, but combine practice with withdrawal in the AI-supported group; Phase-B group contrasts therefore assess the additional unaided benefit of prior assistance. Against the do-nothing default, the groups meet the registered equivalence threshold (between-person $d = -0.05$; $95\%$ equal-tail CrI $[-0.30, 0.20]$, $P(|d| < .40) = .997$), but that scale is at the ceiling for a third of the sample ($66$ of $200$ participants beat passivity in every unaided month). The pre-specified persistence reference, scoring each month against the participant's own prior decisions, has no ceiling (highest observed rate $.91$), and on it \condai{} participants sit slightly above \condnoai{} unaided ($b = +0.24$, $z = 2.29$, $p = .022$; between-person $g = 0.32$, 95\% CI $[0.04, 0.60]$), consistent with partial procedural carry-over.

\subsection{Exploration}
\subsubsection{The advisor suppressed systematic probing, and that did not explain the collapse}
VOTAT probing is a documented pathway to structural knowledge~\cite{vollmeyer1996impact,greiff2012dynamic}, so its suppression under assistance is a candidate pathway to reduced learning. We score VOTAT from the decision log: a month counts as systematic single-variable variation when exactly one of the twelve controls changes from the prior month \emph{and} that control differs from the one isolated on the prior such month, so repeatedly nudging the same control counts as tinkering rather than probing. A participant's score is the share of their transitions meeting this rule.
Systematic VOTAT in aided months was 0.089 against 0.148 in \condnoai{}, a reduction of two-fifths ($g = -0.55$, 95\% CI $[-0.84, -0.27]$, $t(195.4) = -3.93$, $p < .001$). After withdrawal the difference diminished ($0.177$ vs. $0.196$, $g = -0.12$, $95\%$ CI $[-0.40, +0.16]$, $p = .39$), so suppression was confined to aided months, not a participant trait. However, this mechanism does not explain the outcomes: Phase-A probing does not predict unaided Phase-B control ($r = -.00$, $p = .97$), nor within each group (\condnoai{} $r = +.11$, $p = .27$; \condai{} $r = -.14$, $p = .16$).

\subsubsection{Transfer was limited, with some procedural carry-over}
\condai{}'s Phase~B sits at the level of \condnoai{}'s second run ($0.794$ for \condai{} against $0.796$ for that run) rather than its \emph{first}, and slightly above that first run on the two ceiling-free measures: the persistence rate ($0.447$ for \condai{} against $0.350$ for that first run; $b = +0.42$, $z = 4.05$, $p < .001$, odds ratio $1.52$) and company value ($193$k against $138$k; $t(173) = 3.45$, $p < .001$, $g = +0.49$, $95\%$ CI $[+0.21, +0.77]$). At B1, the mean share of controls within $15\%$ of participants' own A12 settings was higher in \condai{} than in \condnoai{} ($0.49$ vs.\ $0.37$). Neither the company-value nor the do-nothing beat-rate comparison detected a group difference at B1. This resemblance is consistent with retained configurations, but does not distinguish procedural from structural learning.

\subsubsection{Recommendation alteration predicted unaided performance}\label{sec:authorship}
\emph{Participants changed recommendations.} We interpret this behavior as \emph{authorship}, not a direct measure of decision generation. Unlike the adherence index's $15\%$ tolerance, alteration means any difference between recommended and submitted values. At least one control differed in $51.3\%$ of $969$ consultations ($864$ distinct months, months~1--12), with a median of two altered controls. Differences concentrated in monthly flow variables---material order, advertising, shirt price, and maintenance---rather than stock or location settings. Material order changed in $29.4\%$ of consultations versus $0.7\%$ for location.

\emph{Alteration was associated with performance during assistance and after withdrawal.} Comparing each submitted decision with the month's final recommendation avoids repeating monthly outcomes across consultations. Altered months beat the do-nothing counterfactual less often, although the binary contrast was unresolved ($87.6\%$ vs. $90.8\%$; $b = -0.44$ log-odds, $z = -1.72$, $p = .085$). More altered controls were associated with lower success ($b = -0.23$ per control, $z = -4.03$, $p < .001$). Alteration frequency in Phase~A predicts unaided Phase-B beat rate ($r = +.36$, $p < .001$), and the association survives adjustment for two measured covariates: Phase-A performance and pre-manipulation exploration-phase ability only slightly reduce it ($b = +0.27 \to +0.24 \to +0.23$; $t(95) = 2.95$, $p = .004$; partial $r = +.29$). A month-level model of the same account agrees ($b = +0.77$ per \emph{SD} of authorship share, $z = 4.11$, $p < .001$).  Participants chose how much to author, so an unmeasured disposition to engage may explain both, and the generation effect~\cite{slamecka1978generation} is an account of this pattern, not a demonstration.

\subsubsection{Who was lifted, and who paid at withdrawal}\label{sec:lifted}
In Phase~B, no between-group difference in end value was detected ($180$k \condnoai{} vs.\ $193$k \condai{}; $p = .55$, $g = +0.12$, $95\%$ CI $[-0.16, +0.40]$), bankruptcies even out, and the A$\to$B change in value retained versus passivity reverses: \condai{} participants fall from A to B where \condnoai{} participants gain ($g = -0.80$, $95\%$ CI $[-1.08, -0.51]$, $p < .001$; the Phase-A advantage over passivity is $g = +0.94$, $95\%$ CI $[+0.65, +1.23]$). Before any advisor appeared, every participant played a six-month exploration phase alone. The value they left it at measures their own skill at the task, uncontaminated by the manipulation. Unaided, it predicts how Phase~A goes ($r = +.50$, $95\%$ CI $[+.33, +.63]$). Under assistance, the same relationship is less than half as strong ($r = +.23$, $[+.03, +.41]$; difference $p = .030$). The advisor lifted the weakest most, with two issues that our study design cannot remove. Assistance compressed outcome variability roughly threefold, so part of the attenuation is range restriction rather than a flattened ability gradient. Separately, a correlation between baseline rank and an A$\to$B \emph{change} score carries regression to the mean by construction. The observed \emph{sign reversal} warrants caution for the same reasons. In \condnoai{} those who began strongest gain least across the reset ($r = -.30$), whereas in \condai{} they gain most ($r = +.30$), so the weakest-baseline \condai{} participants lose most (difference $p < .001$; see \autoref{tab:stratification}). Still, weak-baseline \condai{} participants do not end unaided below their \condnoai{} counterparts (both scales $p > .05$).

\begin{figure}[!htbp]
\centering
\includegraphics[width=\textwidth]{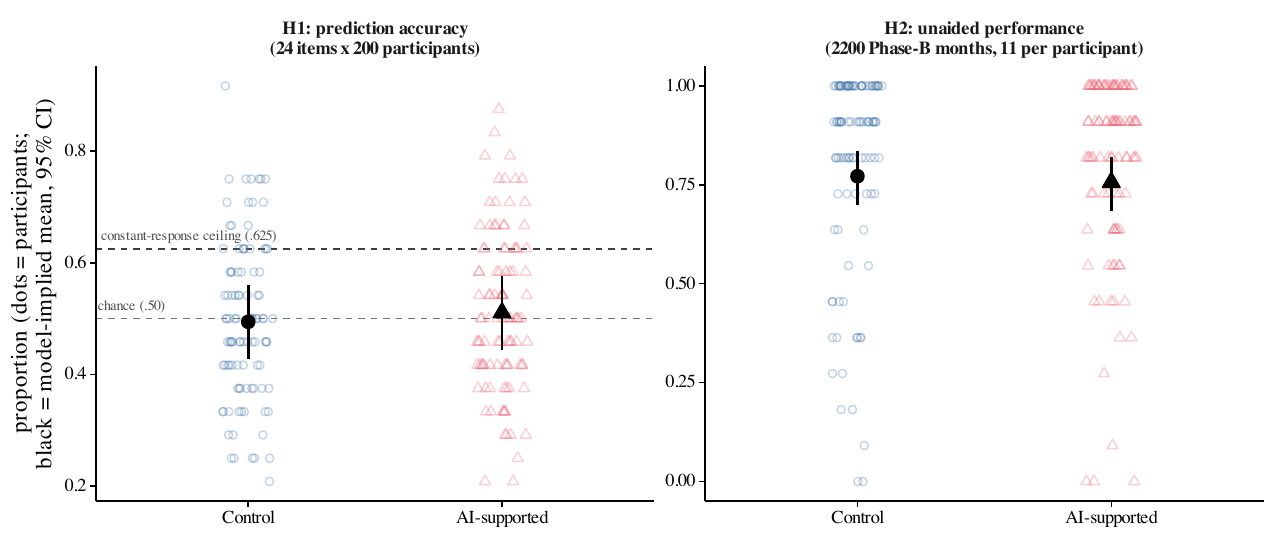}
\caption{\textbf{The aided performance advantage vanishes at withdrawal, while no difference in prediction accuracy is detectable on this battery.} The confirmatory dependent variables by condition: H1 post-Phase-A prediction accuracy and the H2 Phase-B state-conditional rate. \textbf{The two panels do not share a denominator}: H1 is $24$ items scored within each of $200$ participants, H2 is $2{,}200$ participant-months ($11$ scored Phase-B months each), so the panels' interval widths are not comparable to one another. Dots are participants (jittered), encoded by both colour and shape. The black point and bar are the \emph{model-implied} marginal mean and its $95\%$ interval, taken from the same mixed models that carry the confirmatory tests ($\text{correct}\sim\text{cond}+(1|\text{pid})+(1|\text{item})$; $\text{beat}\sim\text{cond}+(1|\text{pid})$) rather than from a raw across-participant CI, so the interval reflects the item and participant variance the tests actually model. On H2 the model-implied marginal ($.77$, $.76$) sits slightly below the observed proportion ($.80$ in both groups) because the participant-level variance is large ($SD = 1.75$ on the logit scale); the groups are indistinguishable on either. Dashed lines on the H1 panel mark chance ($.50$) and the ceiling a constant ``lower'' response would reach against the unbalanced key ($.625$).}
\Description{Two panels, one per confirmatory dependent variable, showing per-participant dots by condition (distinguished by color and shape) with the model-implied mean and its 95 percent interval, plus dashed reference lines at chance and the constant-response ceiling on the first panel. The \condai{} group sits negligibly higher on prediction accuracy, and the two groups are nearly identical on the Phase-B state-conditional rate, with large overlap between conditions.}
\label{fig:hyp_summary}
\end{figure}

\subsubsection{Elevated self-assessments.} (\autoref{fig:understanding_traj}) Across all 200 participants, with the advisor available, \condai{} participants reported higher scores on all subjective measures and lower mental effort (confidence $56.6$ vs. $48.1$, $g = +0.44$, $95\%$ CI $[+0.16, +0.72]$, $p = .002$; understanding $4.75$ vs. $4.27$, $g = +0.36$, $95\%$ CI $[+0.08, +0.64]$, $p = .012$; mental effort $g = -0.45$, $95\%$ CI $[-0.73, -0.17]$, $p = .002$). Removing the advisor eliminated these advantages (phase$\times$condition: confidence $-8.1$ points, $g = -0.68$, $95\%$ CI $[-0.97, -0.40]$, $p < .001$; understanding $-0.33$, $g = -0.45$, $95\%$ CI $[-0.73, -0.17]$, $p = .002$; effort increased only for \condai{} users, $g = +0.50$, $95\%$ CI $[+0.22, +0.78]$, $p < .001$). Across-phase self-assessments were weakly related to prediction accuracy: understanding correlated at $r = .00$ (\condnoai{}) and $r = +.12$ (\condai{}); confidence at $r = -.03$ and $r = +.03$. All $95\%$ intervals included zero, and Bayesian analyses gave moderate evidence for the null (\autoref{tab:bayesnull_subj} in \autoref{app:bayes-nulls}); correlations spanned $-.03$ to $+.13$ across all four self-ratings. Nor did the advisor sharpen confidence at the item level: metacognitive sensitivity, the gap between confidence on correct and incorrect items, did not differ between groups ($g = +0.11$, Bayes factor for the null over the alternative $\mathrm{BF}_{01} = 4.8$), and neither did its non-parametric counterpart, the type-2 area under the receiver operating characteristic curve ($+0.14$, $\mathrm{BF}_{01} = 3.8$). Assistance raised how confident participants felt without improving what that confidence discriminated.

\subsubsection{They could not see the shortfall in their own results.} The self-ratings above are
subjective scales. Participants also estimated their own end-of-phase company value, a measure with a
correct answer. Both estimates were supplied by $193$ participants. In Phase~A the groups earned $144.0$k
(\condnoai{}) and $230.8$k (\condai{}) and estimated $281.5$k and $280.0$k: an $87$k gap in what they
achieved against a $1.5$k gap in what they believed, with the estimate essentially uncorrelated with
performance in either group ($r = +.13$, $p = .21$; $r = -.02$, $p = .87$). At withdrawal the groups'
outcomes differed by $74.6$k (\condnoai{} $+36.9$k, \condai{} $-37.7$k; $t(170.5) = 5.16$, $p < .001$)
while their estimates moved in parallel, differing by $10.1$k ($t(190.2) = -0.72$, $p = .47$).
\condnoai{} overestimation accordingly fell from $+137.5$k to $+64.2$k as those firms improved
($t(94) = -5.02$, $p < .001$), whereas \condai{} overestimation did not move ($+49.2$k to $+60.7$k,
$t(97) = 1.01$, $p = .31$) although their firms lost $37.7$k, and $64.3\%$ of them placed their unaided
outcome at or above what they had achieved with the advisor. The blindness is to magnitude rather than
direction: within \condai{} the change in estimate did track the change in outcome ($r = +.29$,
$95\%$ CI $[+.09, +.46]$, $p = .004$), where in \condnoai{} it did not ($r = +.07$, $p = .47$).

\begin{figure}[!htbp]
\centering
\includegraphics[width=\textwidth]{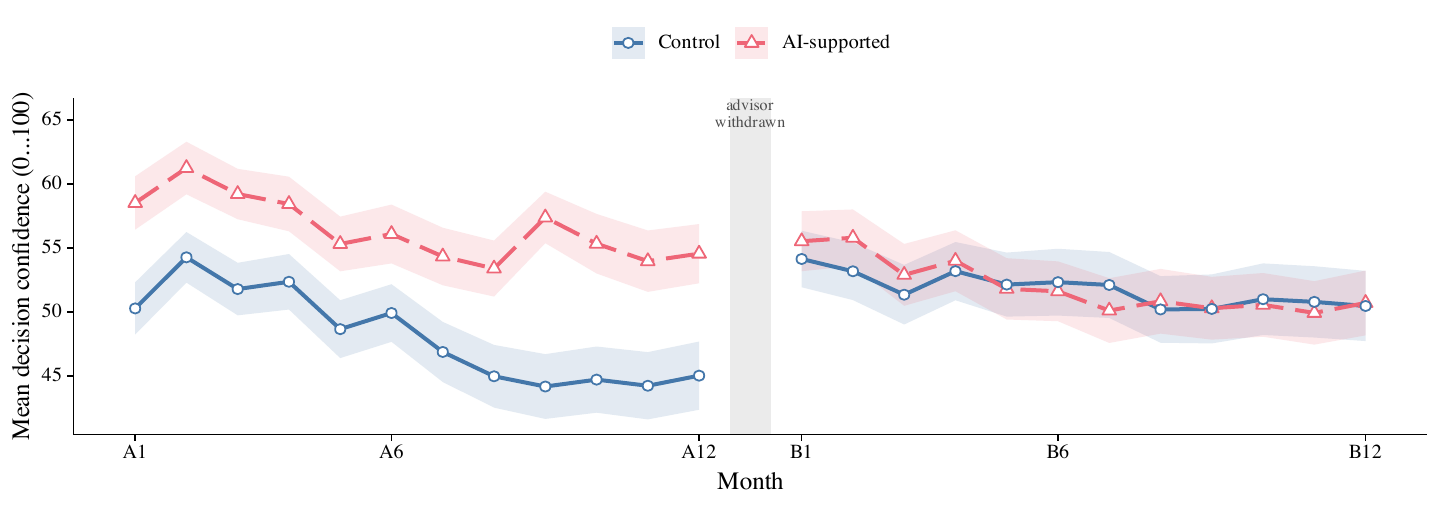}
\caption{\textbf{Confidence anchored to the system tracks its visible state; confidence anchored to the advisor leaves with it (E4).} Mean decision confidence per round across both phases for the $199$ complete reflection sequences (ribbons are $\pm$1~SEM across participants); the grey band marks the between-phase reset and, for \condai{}, the advisor's withdrawal. \condnoai{} confidence declines as the firms lose value, rebounds sharply at the identical Phase-B reset, and holds through Phase~B; \condai{} participants show no such rebound and decline across the unaided months, at performance equal to \condnoai{}.}
\Description{A line plot of mean decision confidence per round over twenty-four rounds spanning both phases, with a grey band at the phase boundary labeled advisor withdrawn. The \condai{} line sits above the \condnoai{} line throughout Phase A; after the boundary, the \condnoai{} line rises to meet it, and the two converge.}
\label{fig:conf_offload}
\end{figure}

\begin{figure}[!htbp]
\centering
\includegraphics[width=\textwidth]{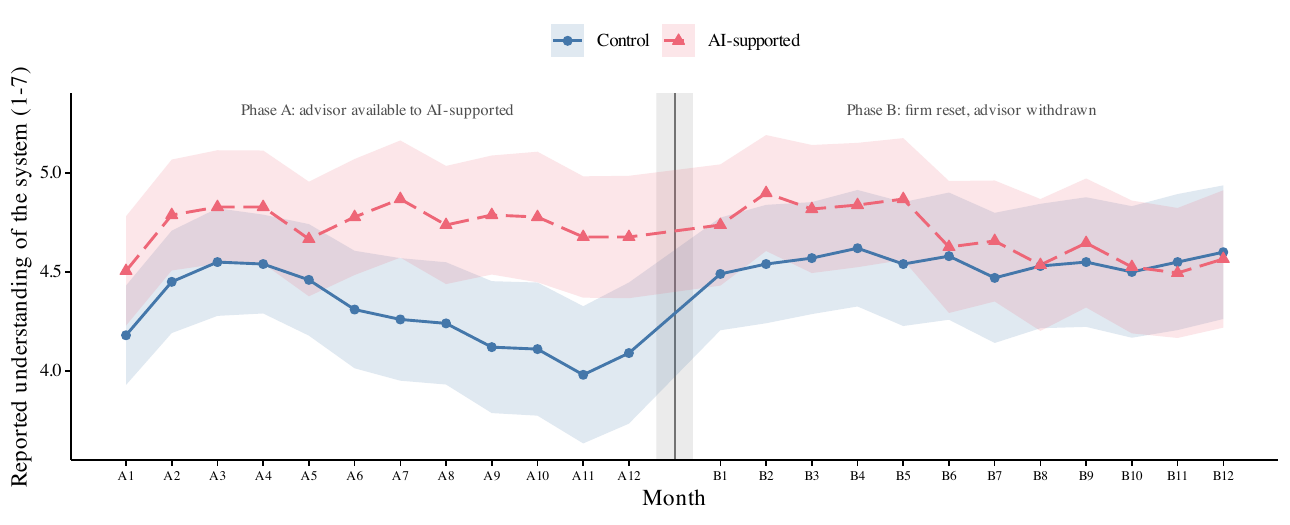}
\caption{\textbf{The felt understanding the advisor supplied is not retained: the gap between
groups closes at withdrawal, and it closes because \condnoai{} rises.} Reported understanding of
the system (1--7), by month and group, for $199$ complete reflection sequences. Ribbons are $95\%$ confidence intervals
across participants, and the grey band marks the reset between phases. While the advisor is available
\condai{} participants report understanding the system better than \condnoai{} participants do ($4.75$ vs.\ $4.27$
averaged over Phase~A). Across the reset itself, the last aided month to the first unaided
one, \condnoai{} participants rise while \condai{} participants barely move, the firm being new and its early months
easy ($+0.40$ vs.\ $+0.06$ points). That asymmetry is reliable ($-0.34$, $t(185) = -2.22$,
$p = .028$, $g = -0.31$, $95\%$ CI $[-0.59, -0.03]$) and larger on confidence ($-8.14$ points, $t(192) = -2.94$, $p = .004$,
$g = -0.41$, $95\%$ CI $[-0.70, -0.13]$; \autoref{fig:conf_offload}). Contrasting whole-phase means rather than the reset
alone gives the same asymmetry more strongly (understanding $-0.33$, $t(188) = -3.18$,
$p = .002$, $g = -0.45$, $95\%$ CI $[-0.73, -0.17]$; confidence $-8.13$, $t(196) = -4.84$, $p < .001$, $g = -0.68$, $95\%$ CI $[-0.97, -0.40]$), and the group gap narrows from $+0.47$ in Phase~A to $+0.14$ in Phase~B.
What the figure shows is
that the aided group's advantage in \emph{felt} understanding does not survive the advisor, on a
measure weakly related to prediction accuracy: within this study, reported
across-phase understanding correlates with prediction accuracy at $r = .00$ (\condnoai{}) and $r = +.12$
(\condai{}), both intervals spanning zero (\autoref{tab:bayesnull_subj}).}
\Description{A line plot of reported understanding across twenty-four months, twelve aided
Phase A months and twelve unaided Phase B months separated by a grey band at the reset. The \condai{} line sits above the \condnoai{} line throughout
Phase A. At the phase boundary both lines rise, the \condnoai{} line rising further, and the two
run close together for the remainder of Phase B.}
\label{fig:understanding_traj}
\end{figure}

Confidence was anchored to the advisor, not to participants' own understanding. Consultation occurred in 73\% of aided Phase-A months, which showed higher confidence than the same participant's unconsulted months ($58.1$ vs. $51.6$ on a $0$--$100$ scale). In a linear mixed model over participant-months (confidence $\sim$ consulted $+$ beat $+ (1\|\text{participant})$), consulting the advisor increased confidence by $+2.8$ points ($t(1037) = 2.66$, $p = .008$), while beating the counterfactual decreased it by $3.4$ points ($t(1000) = -2.89$, $p = .004$; $1{,}089$ aided Phase-A months from $99$ participants). Both effects survived participant fixed effects, the strictest within-person estimator ($+2.5$, $t(988) = 2.39$, $p = .017$; $-3.5$, $t(988) = -3.03$, $p = .003$). The negative effect was not due to the firm's visible trajectory: beating the counterfactual was more likely in declining states, but controlling for month-over-month value changes strengthened the effect (beat $-4.4$, $t(998.27) = -3.38$, $p < .001$; consulted $+2.5$, $p = .017$; value-change term $p = .07$). The negative association did not appear between participants, where confidence and performance were unrelated ($b = +0.5$, $p = .807$). It emerged only within participants ($-4.3$ points, $t(68) = -2.86$, $p = .006$, $d_z = -0.34$), suggesting individuals felt less confident in months when their own decisions worked better. Month-by-month traces are shown in \autoref{fig:conf_offload}. The advisor mainly changed the source of decision-making. The observed transfer is consistent with procedural carry-over, but does not exclude conceptual learning.

\begin{table}[t]
\caption{\textbf{The measures tabulated here differ during assistance, but show no detected
between-group differences after withdrawal.} Summary of the exploratory measurement families
(E1--E5), all uncorrected. Hedges' $g$ is \condai{} $-$ \condnoai{}; ``Phase A'' is the aided
contrast, ``Phase B'' the unaided one. \textbf{Tests.} Strategy and subjective measures are
Welch two-sample $t$-tests on participant-level means. Decision latency is a Mann--Whitney $U$
test on per-participant medians (the distribution is right-skewed). End-of-phase value is a
Welch $t$-test on \$k. Prediction accuracy and metacognitive sensitivity are elicited once,
after Phase~A, so they have no phase contrast. Significance is flagged as
$^{*}p<.05$, $^{**}p<.01$, $^{***}p<.001$; unflagged cells are non-significant, and every
Phase-B cell is. Full statistics per family are in the materials release.}
\Description{A table listing exploratory measurement families with their effect sizes in Phase A
and Phase B. Every Phase-A effect shown is significant. No Phase-B effect is.}
\label{tab:exploratory_summary}
\footnotesize
\setlength{\tabcolsep}{5pt}
\begin{tabular}{@{}p{0.28\linewidth}p{0.28\linewidth}rr@{}}
\toprule
\textbf{Family} & \textbf{Measure} & \textbf{Phase A $g$} & \textbf{Phase B $g$} \\
\midrule
Outcome (E1) & End-of-phase value & $+0.94$\rlap{$^{***}$} & $+0.12$ \\
Process (E3) & Intervention rate & $+0.49$\rlap{$^{***}$} & $-0.07$ \\
 & Decision latency & $+0.74$\rlap{$^{***}$} & $-0.07$ \\
Subjective (E4) & Decision confidence & $+0.44$\rlap{$^{**}$} & $+0.01$ \\
 & Mental effort & $-0.45$\rlap{$^{**}$} & $-0.12$ \\
Self-assessment (E5) & Value-estimate error & $-0.66$\rlap{$^{***}$} & $-0.03$ \\
Mental model & Prediction accuracy & \multicolumn{2}{c}{$+0.12$ (once, after Phase~A)} \\
 & Metacognitive sensitivity & \multicolumn{2}{c}{$+0.11$} \\
\bottomrule
\end{tabular}
\end{table}

\subsection{How participants described using the advisor}\label{sec:s1-rationales}
Our exploratory reflexive thematic analysis~\cite{braun2019reflecting} focused on the $100$ \condai{} participants' Phase~A rationales to examine reliance while advice was available. One author coded the accounts in ATLAS.ti. Reading the coded extracts through the question of who shaped decisions, we developed four themes from five code groups, combining evaluations of advice with delegation.

\paragraph{Testing whose judgment works, and returning after losses.}
Participants described reconsidering whose judgment to follow as outcomes unfolded, reflected in codes for probing, comparing judgments, and returning after losses. P018 explained, ``I did listen to AI this time as I've not made the money i expected with my own decisions'' (month~3). P077 put it bluntly, ``i ruined it, back to AI'' (month~7). Yet losses could also prompt rejection. P018 later wrote, ``I chose not to ask AI this month as i lost money and ended up with no materials'' (month~7). Setbacks thus prompted reassessment, not uniformly greater reliance.

\paragraph{Accepting advice as a basis for action.}
Alongside outcomes, the apparent sense of a recommendation provided a reason to follow it. P016 wrote, ``The AI's recommendation made sense, so I followed it'' (month~2). Codes for trust, agreement, skepticism, following advice, and control captured these judgments and the delegation they could support. P070 described ``giving ai full control this month'' (month~5). Participants could therefore position the advisor as the decision's principal author, even though they still submitted it themselves.

\paragraph{Acceptance left room for selective authorship.}
Delegation was not necessarily wholesale. Codes for mixing, partial use, and overwriting advice captured decisions assembled from both sources. P058 wrote, ``Adjusted the AI on materials and accepted AI on advertising.'' (month~1). Here, acceptance and intervention coexisted within one decision, with the participant retaining authorship over some settings while adopting others.

\paragraph{Advice directed attention to the system.}
Advice could also change what participants noticed, rather than only which settings they chose. Codes for insight, feedback, and identifying problems captured this role. P043 wrote, ``agreed with the AI, i had initially missed the downturn in customer interest'' (month~7). This account describes recognizing a previously overlooked condition, but leaves open whether that recognition would remain available without the advisor.

\section{Study~2: Testing for Structural Knowledge}\label{sec:study2}
Study~1 did not measure its intended construct.  Participants were not able to forecast the three-month consequence of a decision above chance, and the scores did not predict later ability to manage the firm independently. The likely reason is the \emph{level} of knowledge measured. In the Tailorshop, forecasting an output requires simulating the system forward, which research has shown individuals find difficult~\cite{danner2011measuring}. Effective control is instead predicted by \emph{structural} knowledge, an understanding of which variables are linked and in which direction, without simulation~\cite{funke2001dynamic} (see also~\cite{schoppek2002examples}). Study~2 therefore presents a new battery measuring structural knowledge, so that H1 can be tested on an instrument able to register it.

\subsection{Design, participants, and power}\label{sec:s2-power}
Study~2 differs from Study~1 as follows. It tests H1 only and ends after the aided phase and knowledge battery, without withdrawal (\autoref{fig:s2flow}). Phase~A ran $36$ rather than $12$ months to allow more learning, and a structural-knowledge battery replaces the prediction task. To shorten the session, we dropped trust, strategy, and manipulation checks, retaining demographics, NCS-6, and CRT. Study~1 participants could not take part in Study~2.
Study~2 was preregistered on AsPredicted (\#304{,}056, \url{https://aspredicted.org/hf3qz8.pdf}). Participants were paid GBP~8 per hour via Prolific, using the same screener as Study~1. The registered target of $100$ completers per group provided $80\%$ power to detect a difference of $4.5$--$5.8$ accuracy points (see \autoref{app:power-s2}).
The groups are demographically comparable on age, gender, education and prior AI experience (all $p \geq .14$; ages $18$--$74$).\footnote{Denominators $101$ and $96$; one \condai{} participant omitted the form.}

\emph{Sample.} \condnoai{} $n = 101$, \condai{} $n = 97$ after the registered exclusions ($n = 198$). Of the $202$ completers, four were excluded: one for failing the attention check, one for exceeding the registered $200$-minute session cap, one for non-engagement, and one for a median decision latency under $15$ seconds. No honeypot was triggered (\autoref{app:deviations}). The \condai{} group therefore ended with three fewer participants than the registered quota of $100$.

\subsection{Structural-knowledge battery}

The item formats come from Kersting and Süẞ content-valid knowledge diagnostic for the Tailorshop~\cite{kersting1995kontentvalide}, which measures structural knowledge by having participants recognize relations among named variables. We keep their formats and replace their content with our engine's variables. Items ask whether a relation exists, which variable drives which, and with what sign.
The $41$ engine-keyed recognition items cover \emph{direct relations} (\textsc{iia}, 20 items with six response options), \emph{mediated chains} (\textsc{iib}, 14 items with four options), and \emph{connectedness} (\textsc{ia}, 7 items with four options). Option order was randomized per participant and item. The item bank supplies distractors, difficulty classifications, and keys. Overall reliability was KR-20 $= .66$ and $\omega_t = .76$. Facet KR-20 values were $.15$, $.67$, and $.37$ for \textsc{ia}, \textsc{iia}, and \textsc{iib}, respectively. 

\subsection{Hypothesis and scoring}\label{sec:s2-analysis}
In Study~2 we test one confirmatory hypothesis, \textbf{H1}: whether prior advisor availability affects the accuracy of the user's mental model of the Tailorshop system. Study~2 retains Study~1's 2PL specification and $\pm 0.40$ ROPE (\autoref{sec:power}), but registers latent ability $\theta$, rather than prediction accuracy, as the primary outcome. 
Chance performance is $.209$, the facet-weighted average of the per-facet guessing rates ($7$ four-option, $20$ six-option and $14$ four-option items: $(7/4 + 20/6 + 14/4)/41$). The \condcomp{} scored $.670$ on the same battery in participant format.

\section{Study~2: Results}\label{sec:s2-results}

\subsection{A small registered knowledge advantage for \condai{}}
\begin{figure}[!htbp]
\centering
\includegraphics[width=\textwidth]{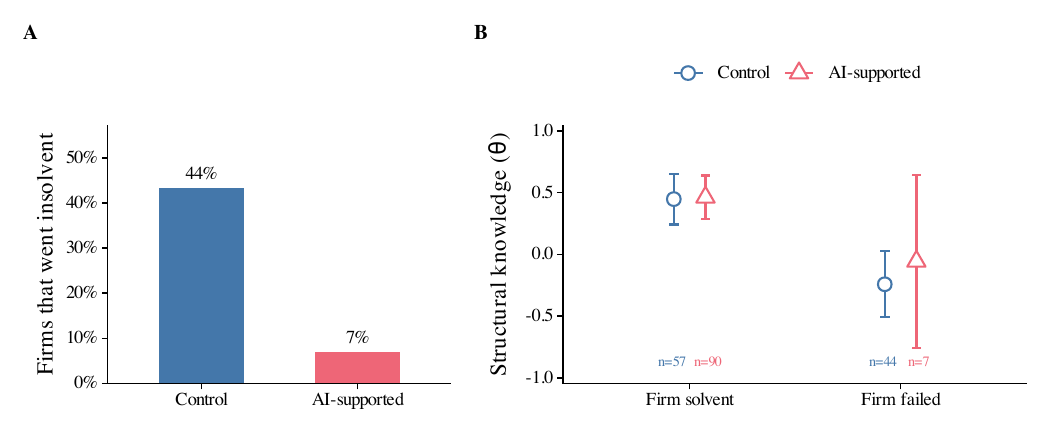}
\caption{\textbf{Study~2's small knowledge advantage is largely associated with differential solvency.}
(A)~\condnoai{} firms failed $44\%$ of the time against $7\%$ under assistance.
(B)~Structural knowledge $\theta$ (registered 2PL; mean $\pm 95\%$ CI) by group within each
solvency stratum. No within-stratum group difference was detected. Adjusting posterior-mean scores for end-of-phase solvency reduces their group contrast from $+0.28$ to $+0.04$, distinct from the registered model contrast of $+0.42$ (\autoref{sec:s2-results}). This post-treatment decomposition is not causal.}
\Description{Two panels. The left bar chart shows 44 percent of \condnoai{} firms went insolvent
against 7 percent of \condai{} firms. The right panel plots structural knowledge for each group
separately among solvent and failed firms. Group means are close within each group,
with solvent firms above failed ones.}
\label{fig:survivorship}
\end{figure}

\emph{Registered confirmatory analysis (latent $\theta$, 2PL, two-sided).} We find a small advantage for \condai{} (\autoref{fig:survivorship}). The registered model contrast puts \condai{} ahead by $+0.42$ ($95\%$ CrI $[+0.11, +0.73]$, $P(\le 0) = .005$).\footnote{Exploratory refits with the original and alternative discrimination priors retained a positive contrast. Specifications and intervals appear in \autoref{app:robustness-details}.} The interval supports a positive effect, but only $46\%$ of the posterior lies inside the $\pm 0.40$ ROPE. Thus, the direction favors \condai{}, while practical equivalence remains unresolved (\autoref{tab:bayesnull_conf}).

This advantage ran contrary to our hypothesis. It was associated with differential solvency. $44\%$ of \condnoai{} firms failed against $7\%$ of \condai{}, and solvent participants scored higher (\autoref{fig:survivorship}). Adjusting the posterior-mean scores for endpoint solvency reduces their group contrast to $+0.04$ ($t(195) = 0.30$, $p = .763$). Among solvent participants, the gap is $+0.01$ ($t(125) = 0.11$, $p = .914$). Note, however, that this post-treatment decomposition is not causal. This analysis cannot separate learning enabled by solvency from pre-existing knowledge associated with it.

Pooling the between-group effects on proportion correct across Study~1 and Study~2 gives $g = +0.14$ ($95\%$ CI $[-0.05, +0.34]$, $p = .15$) with no detectable heterogeneity ($Q(1) = 0.04$, $p = .84$). This pooled estimate combines different instruments and does not resolve the Study~2 knowledge benefit.\footnote{Item-level accuracy specification: $+0.106$ log-odds ($95\%$ CI $[-0.070, +0.282]$), $\chi^2(1) = 1.38$, $p = .241$; $d = +0.06$.}

\paragraph{Interpreting the scores.} Mean accuracy was $.396$ ($SD = .113$), above chance ($.209$, $t(197) = 23.14$, $p < .001$, $d = +1.64$). Scores correlated with the rising-months rate ($r = +.252$, $95\%$ CI $[+.117, +.379]$, $t(196) = 3.65$, $p < .001$) and final company value ($r = +.235$, $p < .001$), supporting their relevance to task performance. Response-strategy and scoring checks are reported in the appendix.

\subsection{Performance separated enormously}
\emph{Performance}. The rising-months rate is $.342$ in \condnoai{} versus $.495$ in \condai{} ($t(196) = 3.61$, $p < .001$, $g = +0.51$, $95\%$ CI $[+0.23, +0.79]$). Insolvency is $44\%$ versus $7\%$ (Fisher's exact test, $p < .001$, odds ratio $9.81$, $95\%$ CI $[4.02, 27.60]$). Performance improved substantially. The knowledge effect was positive under the registered 2PL but unresolved under the accuracy-scale analysis.

\subsection{Knowledge predicted performance mainly in the unaided group}
Posterior-mean $\theta$ correlates with the rising-months rate at $r = +.388$ in \condnoai{} ($p < .001$, $n = 101$) versus $r = +.162$ in \condai{} ($p = .113$, $n = 97$). A month-level model ($\text{rose} \sim \theta \times \text{condition} + (1\,|\,\text{participant})$, with $\theta$ standardized) supports this difference ($b = -0.73$, $z = -2.61$, $p = .009$). The participant-level interaction is less precise ($b = -0.07$, $t(194) = -1.64$, $p = .103$). Restricting the analysis to solvent firms reduces the performance range and yields no detected interaction. Full sensitivity estimates are in \autoref{app:robustness-details}. Knowledge was measured after performance, so the association does not establish a causal effect of knowledge or assistance on their relationship.

\subsection{What participants authored predicts what they knew}
We can replicate the authorship pattern of Study 1. Across $2{,}185$ consultations ($2{,}044$ consulted months) from $95$ aided participants, the submitted decision differed from the recommendation in $31.8\%$ of consultations, a median of $2$ controls when it was, and $10$ participants never altered anything, about $62\%$ of Study~1's alteration rate, over a phase three times as long. What it predicts here is knowledge rather than retention, because Study~2 has no withdrawal. How often a participant altered the recommendation correlates with latent structural knowledge ($r = +.330$, $p = .001$; proportion correct $r = +.216$, $p = .035$) and does \emph{not} correlate with their performance ($r = +.058$, $p = .58$). Entered together, authorship predicts $\theta$ with performance held constant ($b = +0.89$, $t(92) = 3.31$, $p = .001$). This association survives adjustment for performance, but prior knowledge or engagement could still explain both alteration and knowledge.
In Study~1 the exploration measures do not predict unaided Phase-B control, the only one reaching significance running in the wrong direction. The quantity carrying both results is narrower: how often the participant altered the advisor's recommendations.

\begin{figure}[!htbp]
\centering
\includegraphics[width=\textwidth]{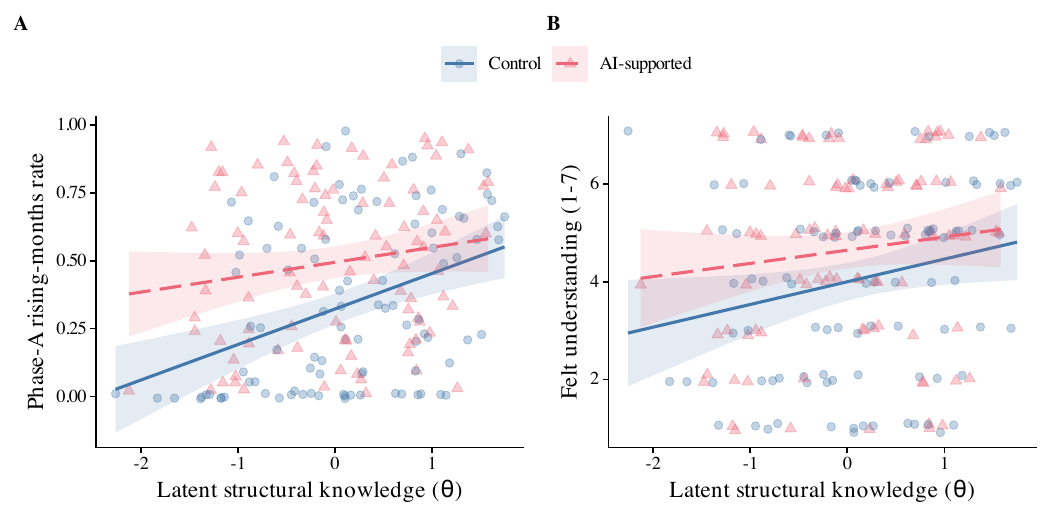}
\caption{\textbf{Structural knowledge is more strongly associated with performance in the unaided group than the aided group; associations with felt understanding are weak.} Latent
ability $\theta$ from the 2PL, registered $n = 198$. Points are participants, jittered
vertically, and lines are least-squares fits with $95\%$ confidence bands.
(A)~Against the Phase-A rising-months rate the slope is clearly positive in \condnoai{} and
near-flat in \condai{}. That difference is tested at the month level in the text. On the
per-person values plotted here it does not reach significance ($t(194) = -1.64$, $p = .103$), so the panel
illustrates the effect rather than carrying it.
(B)~Final-month understanding is modestly associated with $\theta$ ($r = +.21$, $p = .032$ in \condnoai{}; $r = +.13$, $p = .221$ in \condai{}). These final-month ratings differ from the across-month means analyzed in the text.}
\Description{Two scatter plots side by side sharing an x-axis of latent structural knowledge. On
the left the y-axis is the Phase-A rising-months rate and the \condnoai{} point cloud trends
upward more steeply than the \condai{} cloud. On the right the y-axis is final-month felt understanding, with modest positive slopes and substantial overlap between groups.}
\label{fig:s2_criterion}
\label{fig:know_vs_feel}
\end{figure}

\subsection{And it felt like understanding}
\begin{figure}[!htbp]
\centering
\includegraphics[width=\textwidth]{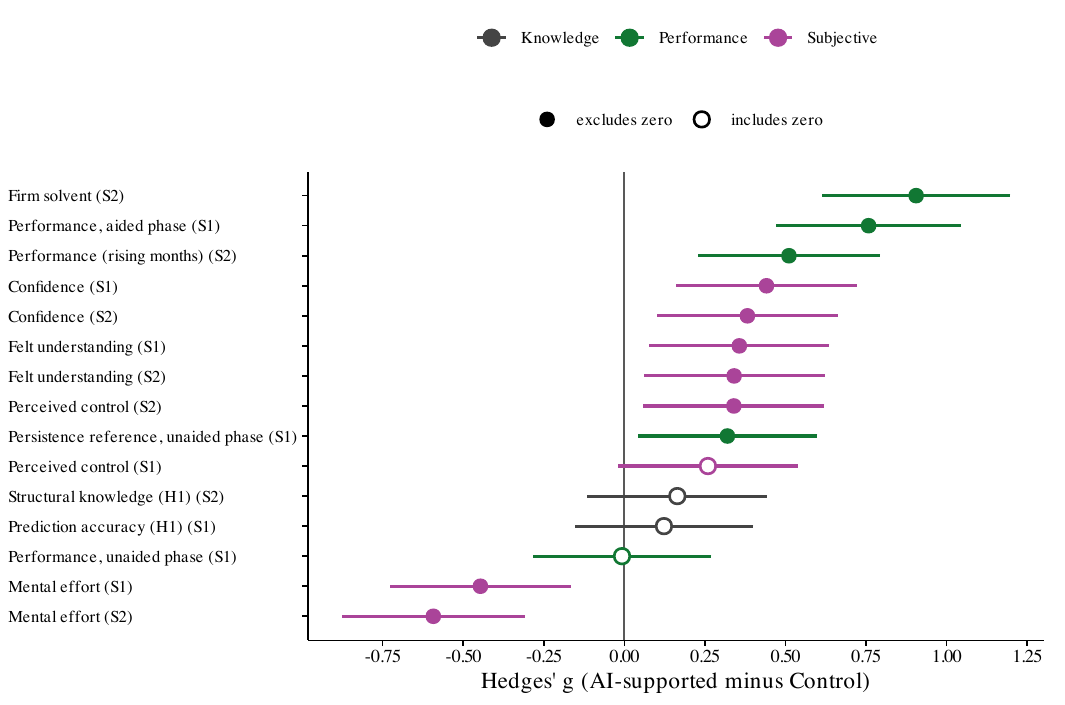}
\caption{\textbf{The same manipulation, in the same sessions, moved performance and self-assessment more than raw knowledge scores.} Every between-group contrast in both studies on one
standardized axis: Hedges' $g$ for \condai{} minus \condnoai{} with $95\%$ confidence intervals,
ordered by magnitude. Filled points exclude zero, open points include it. So that the axis is
comparable, each
contrast is recomputed here as a between-person $g$ on per-participant means, which is not always
the estimator the text uses for the same quantity: in particular the unaided-performance point is the
Phase-B \emph{level} contrast between groups, whereas H2's registered test is the
phase\,$\times$\,condition interaction, which is large and significant
(\autoref{sec:prelim}). The knowledge points are proportion correct, the one scale the two
batteries share. Study~2's registered analysis was latent $\theta$ from a 2PL and is reported in
\autoref{sec:s2-results}. Performance while the advisor is present is the largest effect in either
study. The subjective registers follow at between one-quarter and two-fifths of a standard
deviation, with confidence and felt understanding close in magnitude across the two samples and
perceived control in the same direction but significant only in Study~2. The raw prediction and structural-knowledge contrasts are
statistically unresolved, with point estimates of $+0.12$ and
$+0.16$. Effort runs the other way, aided participants reporting less
of it. Because every contrast is standardized within the same participants and the same sessions,
the figure compares magnitudes. It does not establish knowledge equivalence. Study~1
$n = 200$, Study~2 $n = 198$; per-contrast $n$ varies slightly with instrument completion.}
\Description{A forest plot of fifteen standardized effect sizes, Hedges' g for AI-supported minus
Control, ordered from largest positive at the top to largest negative at the bottom. Firm solvency in Study 2 (about plus 0.91) and aided-phase performance in Study 1 (about plus 0.76) sit near plus
one as filled points, followed by performance on rising months in Study 2 (about plus 0.51). The
confidence, felt understanding and perceived-control contrasts from both studies cluster between
plus 0.26 and plus 0.44, mostly filled, and the unaided-phase persistence-reference contrast from
Study 1 (about plus 0.32) sits filled among them. Near zero, as open points with intervals crossing
it, are the prediction and structural-knowledge contrasts (about plus 0.16 and plus 0.12) and the Phase-B
performance level contrast (about zero). The two mental-effort contrasts sit as filled points
between minus 0.45 and minus 0.59.}
\label{fig:forest}
\end{figure}

\emph{Elevated self-assessments replicate over a horizon three times as long.} Every subjective register separates as it did in Study~1, and by nearly the same amount. \condai{} participants report higher confidence ($55.3$ vs.\ $47.2$, $t(195) = 2.70$, $p = .008$, $g = +0.38$, $95\%$ CI $[+0.10, +0.66]$; Study~1, $56.6$ vs.\ $48.1$, $g = +0.44$, $95\%$ CI $[+0.16, +0.72]$), a higher sense of understanding ($4.64$ vs.\ $4.16$, $t(191) = 2.40$, $p = .017$, $g = +0.34$, $95\%$ CI $[+0.06, +0.62]$; Study~1, $4.75$ vs.\ $4.27$, $g = +0.36$, $95\%$ CI $[+0.08, +0.64]$) and higher perceived control ($4.74$ vs.\ $4.24$, $t(196) = 2.40$, $p = .017$, $g = +0.34$, $95\%$ CI $[+0.06, +0.62]$; the same direction in Study~1 but short of significance, $g = +0.26$, $95\%$ CI $[-0.02, +0.54]$, $p = .067$), all at \emph{lower} effort ($4.62$ vs.\ $5.51$, $t(195) = -4.18$, $p < .001$, $g = -0.59$, $95\%$ CI $[-0.88, -0.31]$; Study~1, $g = -0.45$, $95\%$ CI $[-0.73, -0.17]$). Effort is the largest of the four gaps in both studies. Across two independent samples and an aided phase three times longer, the group means sit within $1.4$ points of each other on confidence and $0.11$ on understanding. Nor is the belief a first impression that fades: the groups begin together and separate steadily, the group$\times$month interaction running $+0.010$ per month for understanding ($t(6928) = 5.03$, $p < .001$) and $+0.15$ for confidence ($t(6928) = 4.62$, $p < .001$) over $7{,}128$ participant-months. Associations with measured knowledge were weak or unresolved in either group ($r$ between $+.04$ and $+.16$, all $p > .10$, for mean understanding and confidence). Study~2 has no withdrawal and cannot test whether the belief survives the advisor's removal. That evidence is Study~1's, where the belief evaporates as it accumulates here. Placing every contrast from both studies on one standardized axis (\autoref{fig:forest}) shows large effects on what participants achieved, moderate ones on what they felt, and smaller, estimator-dependent effects on what they knew.

\section{General Discussion}

Two studies examined the distinction between \emph{augmentation} and \emph{substitution} in AI-supported complex problem solving. Study~1 withdrew the advisor after twelve aided months and measured what participants could still do alone. Study~2 kept the advisor throughout and tested how post-phase knowledge related to performance
(\autoref{sec:study2}).

While the advisor was available, participants receiving assistance ran more functional firms, with substantially less insolvency. They reported feeling more capable and in control, while exerting less effort. In Study~1, however, the large aided performance advantage largely disappeared in the first unaided month.

\subsection{Performance was lent, the knowledge benefit was less clear}
Neither study supported the knowledge cost we predicted (H1). Study~1 detected no group difference, but its battery performed at chance and did not predict unaided performance. It therefore tells us more about the instrument's failure than about whether the groups learned equally. The evidence for a knowledge benefit comes from Study~2, where the registered contrast put \condai{} slightly ahead.

This benefit was closely tied to solvency. Participants whose firms remained solvent knew more about the system, and assistance substantially reduced failure. Staying solvent could preserve opportunities to learn even as receiving recommendations reduced the need to generate decisions. Within solvency strata, no group difference was detected. Because solvency is itself an outcome of assistance, these comparisons unpack the registered contrast rather than explain it away (\autoref{fig:survivorship}).

Study~1's withdrawal test (H2) addressed the other part of capability, what participants could still do alone. The large aided advantage largely disappeared after withdrawal, including among the weakest participants. The persistence reference nevertheless showed a small unaided advantage. Together with Study~2's positive knowledge contrast, this shows why learning and capability lending need not be alternatives. Some benefits remained, but most of the performance advantage depended on the advisor.

\subsection{Capability lending}\label{sec:discussion-construct}
We use \emph{capability lending} to describe this pattern. An assistant improves decisions, outcomes, and felt understanding more than it develops the user's independent capability, so most of the performance benefit depends on continued assistance. The term describes the pattern rather than a distinct mechanism. One possible explanation is that a reliable advisor reduces the immediate return to understanding the system (cf.\ active inference~\cite{friston2015active}).

This connects to complacency and the out-of-the-loop problem~\cite{bainbridge1983ironies,parasuraman2010complacency,wickens2015complacency,onnasch2014human} through the allocation of work. On the levels-of-automation taxonomy~\cite{endsley1999level}, our advisor automated option generation and selection, while participants retained implementation. The concern here extends beyond monitoring imperfect automation to what users retain after frequently following beneficial advice. Berry and Broadbent distinguish successful control from articulable knowledge~\cite{berry1984relationship}. Withdrawal adds the question of whether that control remains possible without assistance. The illusion of competence~\cite{he2023knowing} concerns inflated self-belief that precedes interaction and depresses reliance. Here, self-assessments rose during assistance, while withdrawal revealed how much performance depended on the advisor.
This makes the user's role in generating decisions particularly important. Bu\c{c}inca et al.\ find that contrastive explanations improve \emph{independent} decision skill~\cite{buccinca2025contrastive}. Their explanations are built against the user's likely answer, inviting comparison, whereas our advisor supplied a recommendation for acceptance or modification. Whether assistance preserves the user's generative step may therefore matter alongside reliability and task difficulty.

The lending metaphor concerns the duration of the benefit, not a formal utility calculation or a claim that the pair exceeds its parts. The pair finished Phase~A below the model alone, though above unaided controls. Human involvement may still be warranted for accountability, context the model cannot see, and continuity when it is unavailable. These roles can justify collaboration without superior task performance, but human presence alone does not establish that they are fulfilled. Withdrawal tests assess one of these requirements, the person's ability to continue alone.

\subsection{A bypass, not a deficit}\label{sec:discussion-explanation}

Reliance stayed high in both studies, yet participants also altered recommendations. Changes clustered in the four monthly flow variables of material order, advertising, shirt price and maintenance, while outlets, worker and machine stock, and location differed less often from recommendations. Study~2 showed the same pattern over three times as many months.
The pattern is consistent with intransparency in Dörner's sense~\cite{dorner2017complex,funke2001dynamic}. Flow variables can have visible same-month effects. Structural variables work through delays and couplings that may require tracking changes over months. Participants kept more of the former and ceded more of the latter, though alteration rates alone do not establish what they understood.

Confidence data suggest self-assessments were weak indicators of measured knowledge. In Study~1, consulting the advisor increased felt confidence, while successful months seemed, if anything, to reduce it. The ease of carrying out a competent plan was experienced as their own competence~\cite{FERNANDES2026108779}. This is compatible with several accounts, including cognitive offloading~\cite{risko2016cognitive,sparrow2011google} and control without articulable knowledge~\cite{berry1984relationship}. The prediction battery cannot identify the knowledge underlying unaided performance.
What remained may include retained settings. In Study~1, aided participants' first unaided decisions resembled their own final Phase-A settings more than controls' decisions did, but no group difference in first-month performance was detected.
In Study~2, knowledge was more strongly associated with unaided performance, although the moderation depended on analysis level.
Thus, participants appear to have retained settings from the aided phase. Whether they also retained the relations that produced them remains unresolved.

\subsection{Authorship predicted what remained}\label{sec:discussion-authorship}
Those who delegated the least did best without the advisor. How much a \condai{} participant adjusted the advisor's suggestions predicts unaided performance in Study~1 after withdrawal, controlling for aided performance and pre-manipulation ability. The association remains unchanged after adjustment for need for cognition and CRT, with neither trait clearly predicting authorship. In Study~2, which never withdraws the advisor, it predicts structural knowledge.

This is consistent with a generation account in which producing material can improve retention relative to receiving it~\cite{slamecka1978generation}. We hold the account loosely for two reasons. Authoring was chosen rather than assigned, so one disposition could explain both authoring and retention. Controlling for pre-manipulation ability and aided performance narrows that explanation without replacing the experiment needed to settle it. More altered controls were associated with lower same-month success. This observational association does not establish a causal cost of altering recommendations. A direct test would assign generation-before-advice versus recommendation acceptance while holding advice quality constant.
Thus, participants described testing advice, accepting plausible explanations, delegating decisions, selectively changing recommendations, and noticing overlooked conditions. Returning to AI after losses suggests that difficulty could renew delegation, although some losses prompted rejection instead. These accounts make reliance a response to unfolding experience, while leaving open whether selective authorship produced lasting understanding.

\subsection{Assistance is felt as understanding}
Assistance raised confidence, perceived control, and felt understanding while reducing effort. A coherent recommendation may make the task feel understood without requiring users to weigh its trade-offs. After withdrawal, confidence returned to baseline before felt understanding declined. Subjective changes thus lagged behind the loss of the performance advantage. Higher ratings and weak knowledge correlations do not establish individual miscalibration, given the knowledge measures' limitations.
The discrepancy also appeared in estimates of company value. Across withdrawal, the groups' actual outcomes diverged much more than their estimates. Control participants' overestimation fell as their firms recovered, whereas aided participants' overestimation persisted as theirs declined. Self-reports during assistance therefore provide an incomplete account of both independent capability and task outcomes.

\subsection{Implications}
Following van Berkel and Hornb{\ae}k's typology~\cite{vanberkel2023implications}, we raise implications for theory, methodology, design, and practice.
We distinguish reliance calibration, the construct typically measured, from reliance investment, the capability repeated reliance builds. Participants frequently followed usually beneficial advice, yet showed limited additional unaided capability. A user who challenges a mediocre advisor may show lower calibration but retain more capability. Calibration alone cannot assess this investment, raising the longer-term question of how organizations develop expertise while automating its practice.

In-use metrics showed assisted performance gains that largely disappeared at withdrawal. Learning and performance diverge when it matters most~\cite{soderstrom2015learning}, and unaided outcome measures help distinguish them. Because assistance is deployed continuously, evaluations restricted to assisted outcomes cannot establish retained capability. Microworlds record every decision, state, and ground truth, so a task model can be checked against the system, not self-report. Dynamic, opaque, polytelic tasks carry the greatest value and cost of assistance yet remain understudied in HCI~\cite{funke2001dynamic,mirel2004interaction}.

Because authorship predicted retention, consistent with the generation effect~\cite{slamecka1978generation}, a fading advisor could present settings early, then later report state and uncertainty without exact decision values. The correlational authorship--retention link motivates an experimental test. Charging per query might encourage independent problem solving. Whether metering reduces delegation or excludes those most aided is untested. The advisor maintained a plan. Whether users did so is unresolved. User-co-authored plan artifacts could extend guidance on understanding AI~\cite{amershi2019guidelines} to understanding the shared task. Monthly recommendations may favor rule-following over knowledge-based control~\cite{rasmussen1983skills,woods2009rasmussen}, but our measures do not establish that distinction. Ecological interface design points the other way, making work-domain constraints visible instead of supplying the answer.

Assistance most benefited users least able to function without it. Similar assisted outcomes may therefore conceal different capacities to continue alone. AI systems are not always available~\cite{10.1145/3772363.3798382}, making continuity of support---and who controls access to it---particularly relevant. Human-in-the-loop oversight and synergy require more than assisted success~\cite{onnasch2014human,buccinca2021trust,vaccaro2024combinations,lee2004trust}. Assessment should distinguish assisted performance, the ability to challenge advice, and the ability to continue without it.

\subsection{Limitations and Future Work}\label{sec:limitations}

First, \emph{evidence for learning is weaker than evidence for assisted performance.} Both groups scored at chance on Study~1's knowledge test. Study~2's knowledge advantage depended on scoring and was closely tied to solvency. Without pretests, neither study quantifies individual knowledge gains. These limitations restrict claims about learning and accurate self-assessment, not the observed performance contrasts. Validated pre- and post-tests~\cite{greiff2012dynamic} should distinguish understanding system relations from learning useful decision rules~\cite{berry1984relationship,funke2014analysis}.

Second, \emph{the withdrawal test establishes immediate transfer under a reset, not lasting independence.} Both groups restarted the firm, while only \condai{} lost assistance. Their Phase-B contrast therefore assesses the additional unaided benefit of prior assistance in that setting. The primary score's ceiling limits sensitivity, but the persistence reference still shows a small benefit. Together, these measures support limited additional transfer, not a complete absence of learning. Separating reset from withdrawal, varying its announcement, and testing retention later~\cite{soderstrom2015learning} would establish how far the pattern extends.

Third, \emph{the studies do not isolate how assistance affects learning.} Recommendation alterations predicted later outcomes after covariate adjustment, but were self-selected. Near-universal uptake limits dose--response analysis, and knowledge-based regressions omit uncertainty in the scores. Required rationales could prompt reflection, although Study~1's aided participants often gave brief, minimally elaborated accounts (\autoref{sec:s1-rationales}). These issues limit the proposed generation account, rather than the observed contrast between assisted and unaided performance. Randomizing generation-before-advice or fading support~\cite{bastani2024generative}, with and without required rationales, would test that account.

\section{Conclusion}
Across two pre-registered studies of complex problem solving, AI advice improved performance and raised confidence and felt understanding while reducing effort. After withdrawal in Study~1, most of the performance advantage disappeared, although a small benefit remained against the persistence reference. In Study~2, the small registered knowledge advantage was closely tied to remaining solvent.
We call this pattern \emph{capability lending}, with much of the performance benefit depending on continued assistance. Among aided participants, more frequent recommendation alterations predicted better unaided performance in Study~1 and higher structural knowledge in Study~2. These associations motivate testing AI interaction designs that preserve the user's role in generating decisions.
For HAI, evaluating tools that extend human capability means assessing assisted performance, measured knowledge, and independent control separately. Better outcomes with AI do not, by themselves, establish greater capability in the person.

\section*{Use of generative AI}
We used large language models (Claude Opus 4.8 \& 5, Claude Fable, and OpenAI Astra) to write and refactor code and to draft and revise prose. All text and code they produced were checked and verified by the authors. The authors take full responsibility for the content.

\begin{acks}
We thank Esko Evtyukov for developing the study prototype and experiment software, and Pekko Vasantola for detailed comments on the manuscript.

This work was supported by the European Research Council (ERC) under the European Union's Horizon Europe research and innovation programme, AmplifAI (grant agreement No.\ 101217557), and by the Research Council of Finland (decision no.\ 371202, MET-AI). Views and opinions expressed are however those of the author only and do not necessarily reflect those of the European Union or the European Research Council Executive Agency. Neither the European Union nor the granting authority can be held responsible for them.

\smallskip
\noindent\includegraphics[width=6cm]{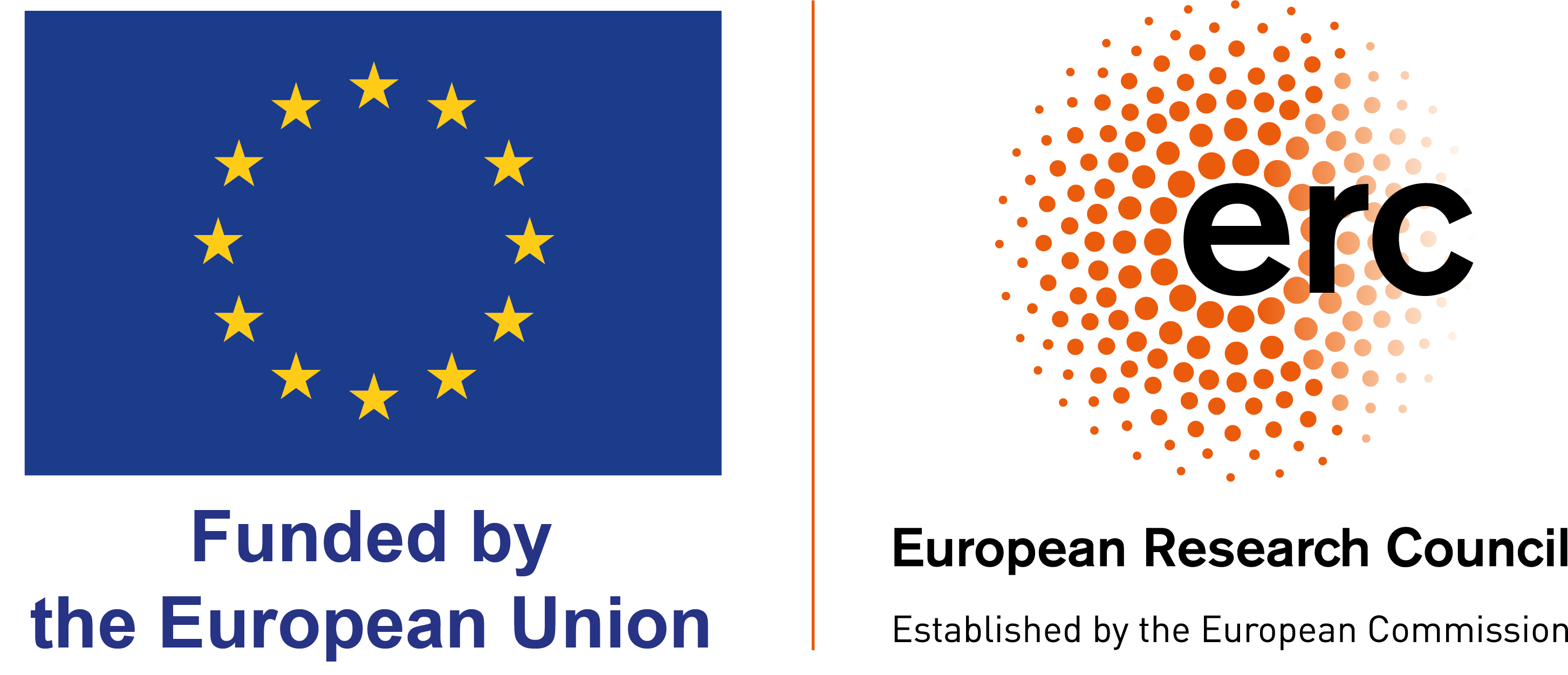}\par
\end{acks}

\FloatBarrier
\bibliographystyle{ACM-Reference-Format}
\bibliography{bibliography}

\appendix

\section{Preregistrations, deviations, and power}
\subsection{Deviations from preregistration}\label{app:deviations}
Six deviations are recorded: five in Study~1 (D1--D5) and one in Study~2 (S1). All but D5 were decisions, and each was decided blind to outcomes. D5 is not a decision but an implementation omission we found after the freeze, and it is reported here on the same terms. 
\paragraph{D1: Manipulation check: reported \emph{use}, not \emph{availability} (2026-07-16).} \emph{Registered:} exclude on availability reports inconsistent with condition. \emph{Deviation:} \condnoai{} participants are excluded for reporting advisor availability and non-zero consultation frequency; the \condai{} criterion is dropped because participants denying availability had nevertheless consulted the advisor. \emph{Effect:} the as-run manipulation screen excludes $12$ \condnoai{} and no \condai{} participants. Across all screening rules, exclusions total $17$ and $4$ respectively; the four \condai{} exclusions comprise two attention failures and two session-limit failures. The comparison screening pipeline excludes $18$ and $11$ overall. Selection bias remains possible. The reported H1 and H2 sensitivities retain their interpretations across the exclusion sets, with H1 estimates from $+0.08$ to $+0.11$ and H2 from $-1.17$ to $-1.23$. Adding fielding batch as a random effect leaves both estimates unchanged. This does not establish robustness of every exploratory result. Per-set estimates are in the materials release.

\paragraph{D2: Allocation of the final batch (2026-07-18).} \emph{Registered:} 1:1 assignment maintained throughout. \emph{Deviation:} the final top-up batch of $46$ was allocated $44{:}2$ toward \condnoai{} to fill the per-group quota, because the revised exclusion rules are structurally asymmetric (the manipulation-check rule can exclude only \condnoai{}). Participants are blind to allocation, procedures remain unchanged, and the stopping rule reads no outcome. \emph{Effect:} no detectable effect. Randomization balance holds on all pre-treatment covariates in the confirmatory sample: age, gender, prior AI experience, need for cognition, cognitive reflection, exploration-phase value, and pre-baseline each have absolute standardized mean differences $\leq .18$ and $p > .20$. The one imbalance, session duration ($68$ vs.\ $77$ minutes, $p = .013$), is post-treatment (consulting an advisor takes time) and is therefore never used as a covariate.

\paragraph{D3: The data selected none of the three registered verdicts.} \emph{Registered:} \emph{substitution}, \emph{augmentation} and \emph{no measurable cost}. \emph{As run:} the outcome fell outside all three, landing in a cell of the full $3\times3$ crossing (\autoref{tab:verdict}) that the registration had not named. \emph{Effect:} none on the analysis. Every definition, contrast, test and threshold is as registered, and the cell labels were fixed in the analysis script before any outcome was read; only the reported label is one the registration did not enumerate.

\paragraph{D4: Both-outcomes analytic sample (post-freeze).} \emph{Registered:} per-test inclusion (H1 $=$ complete battery; H2 $=$ both-phase completer). \emph{Deviation:} the confirmatory sample is restricted to the $200$ participants ($100$ per group) with complete data on \emph{both} primary outcomes, so H1, H2, and every downstream analysis describe one identical sample; this drops one H1-only and ten H2-only participants relative to the registered per-test sets ($101/100$ and $105/105$). \emph{Effect:} the confirmatory estimates are materially unchanged, and the groups stay balanced.

\paragraph{D5: Registered exclusion (iv) was never implemented (found post-freeze).} \emph{Registered:} exclude for technical failure or non-engagement: more than three browser disconnects within Phase~A, an engine state-corruption flag, or a scored phase in which the participant kept the pre-filled decision defaults unchanged for all but at most one month. \emph{As run:} the confirmatory script omits this rung; the omission is ours and we found it only after the freeze, so no participant was tested against it. \emph{Effect:} applying the registered rule to the frozen data excludes exactly one \condnoai{} participant, and no \condai{} participant is affected. H1 then moves from $+0.083$ to $+0.078$ and H2 from $-1.227$ to $-1.226$, with no verdict cell, ROPE decision or direction changed. The paper reports the as-run $200$ throughout. Applying this exclusion slightly reduces the H1 estimate without changing its interpretation.

\paragraph{Study 2 deviations.} One, on the same terms, lettered S1 to keep it distinct from Study~1's D1--D5. Two further Study~2 choices are analytic specifications rather than departures and are recorded as such below.

\paragraph{S1: Recruitment stopped three short of the registered quota.} \emph{Registered:}
recruit to $100$ completers per group, continuing until both groups reach quota. \emph{As run:}
recruitment stopped with \condnoai{} at $101$ and \condai{} at $97$. \emph{Why:} the reason was
timing (we closed the study at the end of July 2026 to prepare the manuscript) rather than outcome-driven. 

\paragraph{Specifications and disclosures, not deviations.} Three choices are recorded here
rather than in the list above, because the registration either left them open, entailed them,
or was not departed from at all.

\emph{Screen-inspection telemetry (Study~1): collected, never analyzed.} Passive client-side logging was added before the first confirmatory session and was not registered as an outcome. On inspection the recorded events proved a poor index of information search, so we withdrew the family rather than report quantities the instrument does not support. No measure in either study touches these data. These unanalyzed raw logs are excluded from the minimal data release.

The remaining two are Study~2's.
\emph{Estimation:} the registration named the crossed-random-effects model without specifying an
approximation. We used Laplace (\texttt{nAGQ = 1}), matching Study~1, and the contrast moves by
$.002$ log-odds under the alternative. \emph{2PL sign alignment:} the 2PL likelihood is invariant
under a joint sign reflection of discriminations and latent scores, so chains land on either
reflection at random and a pooled summary averages across both. We align draws per chain before
summarizing. This is an orientation convention motivated by the likelihood's reflection symmetry: without it the same healthy fit
summarizes to $|a| \approx 0$ for every item and an empirical reliability of $.00$, and the same
correction applies to Study~1's item-response reanalysis. We additionally
withhold a ROPE verdict unless $\hat{R} < 1.1$ and bulk effective sample size (ESS) $\geq 1{,}000$ on the
contrast. That gate is ours, not the registration's, and it changes no reported verdict.

\subsection{Power analysis details}\label{app:power}
Both studies were sized by Monte-Carlo simulation of the confirmatory mixed models
(\autoref{sec:analysis-plan}) rather than from a two-sample summary, over analytic $n$ per group
$\in \{60, 80, 100, 120\}$ with $400$ samples per cell. All nuisance parameters were set a priori
and no pilot data informed the Study~1 power parameters. The simulation scripts, the cell-by-cell power
tables and the ROPE decision-rate simulation are in the materials release. At the planned $n = 100$
per group, H1 reaches $0.84$ power for an $8$-point accuracy gap and is underpowered ($0.49$) for a
$5$-point gap; H2 reaches $.91$ for the registered withdrawal interaction.

\subsection{Study~2 power analysis details}\label{app:power-s2}
The Study~2 target of $100$ completers per group (\autoref{sec:s2-power}) was set by Monte-Carlo
simulation of the registered item-level model. Two features drive it and are worth stating here.
The battery is forced choice with \emph{guessing floors} ($1/6$ on the $20$ direct-relation items,
$1/4$ on the other $21$), which compress the accuracy difference a given latent-ability difference
can produce: at $d = 0.40$ the floors reduce the observable shift from $7.5$ to $3.3$ points. Data
were generated from a three-parameter model carrying those floors and analyzed with the registered
two-parameter model, so the misspecification was simulated rather than assumed away. Item
discrimination is the least certain parameter, so power is reported as a band. Full parameters and
the simulation script are in the materials release.

\section{Method details}\label{app:method-details}
Reference-level detail moved out of the main text for concision. Nothing here alters the registered design.
\paragraph{Recruitment screeners.} Prolific screeners: first language English, current country of residence in UK, approval rate $\geq 95\%$, no participation in any of our prior studies on this platform (custom blocklist). Inclusion criteria measured on the platform itself: age 18--65, no prior participation in business simulation studies, and no prior exposure to the Tailorshop. One participant in each study self-reported an in-study age above this ceiling ($67$ in Study~1, $74$ in Study~2); both were retained.

\paragraph{The 2PL as a measurement model.} A 2PL is a measurement model rather than a scoring rule: it estimates each item's difficulty and its \emph{discrimination}, how sharply the item separates people who know from people who do not, and reports the ability that best explains a person's whole response pattern. A sum score instead gives every item one equal vote, so an item that fails to separate knowers from non-knowers counts as much as one that separates them cleanly, and negatively discriminating items are averaged in rather than shown. The confirmatory model of \autoref{sec:power} is the one-parameter special case, discriminations held equal, which is what makes it equivalent to counting correct answers. See~\cite{burkner2021bayesian} for an introduction, and \texttt{docs/model\_assumptions.md} in the materials release for the full psychometric diagnostics.

\paragraph{Tailorshop state and controls.} The engine separates two things the dashboard shows together. \emph{Decisions} are what the participant sets each month, free to take any value the sliders allow; \emph{state} is what the engine carries between months and mutates through its equations, taking the decisions as inputs. The state the engine carries between months comprises the month index, company value, bank-account cash, shirt sales, customer interest, the raw-material price, raw-material and finished-shirt stock, machine capacity, machine damage, idle production, and worker satisfaction. The twelve monthly decision controls: hiring of the two worker types, the wage rate and worker benefits, advertising spend, retail price, the number of sales outlets and their location, raw-material orders, purchases of the two machine types, and machine maintenance. Demand was a non-linear function of customer interest and retail price, with customer interest itself set each month by advertising, sales outlets and location plus a fixed random term, so month-to-month persistence enters through interest rather than through demand; production capacity a function of workers, machinery count, and machinery condition (which decayed unless maintained); monthly expenses aggregated material orders, salaries, worker benefits, outlet rent, storage, advertising and maintenance; company value aggregated bank-account cash, raw-material and finished-shirt stock at fixed unit values, machines at their capacity-depreciated worth, and sales outlets net of a per-month write-down. The exact parameterisation and initial state will accompany the materials release upon acceptance.

\paragraph{Exclusion operationalisation.} (i)~Failed attention check: any of the four attention/comprehension items wrong after the second try. (ii)~Failed manipulation check: in \condai{}, reporting no AI advisor was available; in \condnoai{}, reporting one was (revised under deviation D1, \autoref{app:deviations}). (iii)~Session length exceeding the registered cap: $150$ minutes in Study~1 and $200$ in Study~2, which registered the longer cap for its longer task. (iv)~Technical failure or non-engagement: browser disconnects $>$\,3 within Phase~A, any state-corruption flag from the engine, or a phase in which the participant kept the pre-filled decision defaults unchanged for all but at most one month (accepting the advisor's suggestions, which change the decisions, counts as engagement). (v)~A triggered attention honeypot (a hidden trap item an engaged participant would not activate). Two pilot-calibrated data-quality screens: (vi)~an implausible company-value forecast above $1{,}000{,}000$ (roughly four times the starting capital and three times the highest value observed in any human or autonomous-AI run) is set to missing; the participant is retained and drops only from analyses using the forecast; (vii)~a scored phase whose \emph{median} per-decision latency falls below 15\,s (engaged pilots produced zero of 220 scored decisions under 15\,s and phase medians of 36\,s or more) excludes the participant.

\begin{figure}[!htbp]
\centering
\resizebox{.9\linewidth}{!}{%
\begin{tikzpicture}[
 font=\sffamily\small, x=1cm, y=-0.78cm,
 shared/.style ={draw, fill=gray!10, align=center, minimum width=10.8cm, minimum height=7.5mm, inner sep=4pt},
 lane/.style ={draw, fill=white, align=center, minimum width=5.3cm, minimum height=7.5mm, inner sep=4pt},
 withai/.style ={draw, fill=cpsai!30, align=center, minimum width=5.3cm, minimum height=7.5mm, inner sep=4pt},
 measure/.style={draw, fill=yellow!40, align=center, minimum width=10.8cm, minimum height=7.5mm, inner sep=4pt, font=\bfseries\small},
 arr/.style ={->, >=Latex, semithick}
]
\node[font=\bfseries] at (-2.65,-0.7) {\condnoai};
\node[font=\bfseries] at ( 2.65,-0.7) {\condai};
\node[shared] (c) at (0,0) {Consent + demographics; attention check};
\node[shared] (s) at (0,1.1) {Cover story + interface walkthrough};
\node[shared] (e) at (0,2.2) {Exploration: 6 months, unaided, unscored (both groups)};
\node[lane] (pn) at (-2.65,3.5) {\textbf{Phase~A} -- months 1--36\\ no advisor};
\node[withai] (pa) at ( 2.65,3.5) {\textbf{Phase~A} -- months 1--36\\ advisor in chat panel};
\node[measure] (b) at (0,4.8) {Structural-knowledge battery (41 engine-keyed items; once, after Phase~A)};
\node[shared] (mc) at (0,5.9) {Debrief (no manipulation check)};
\draw[arr] (c) -- (s); \draw[arr] (s) -- (e);
\draw[arr] (e.south) ++(-2.65,0) -- (pn.north); \draw[arr] (e.south) ++(2.65,0) -- (pa.north);
\draw[arr] (pn.south) -- (b.north -| pn.south); \draw[arr] (pa.south) -- (b.north -| pa.south);
\draw[arr] (b) -- (mc);
\node[font=\footnotesize\itshape, gray!50!black, anchor=west, text width=4.9cm] at (5.7,3.5)
 {Per round: self-report battery as in Study~1; full action logging; NCS-6 and CRT retained; no trust scale, no strategy survey, no manipulation check, no Phase~B.};
\node[font=\footnotesize\itshape, gray!50!black, anchor=west, text width=4.9cm] at (5.7,4.8)
 {Scored server-side against the engine-verified key (\autoref{sec:s2-analysis}).};
\end{tikzpicture}%
}
\caption{\textbf{Study~2 keeps Study~1's aided phase and drops everything downstream of the mental-model probe.} Study~2 flow: shared onboarding and unaided exploration, a single 36-month Phase~A (advisor in \condai{} only), the 41-item structural-knowledge battery, and the closing checks. There is no Phase~B and no withdrawal; the AI-alone benchmark and engine seeds match Study~1.}
\Description{A two-lane flow diagram for Study 2. Both lanes share consent, walkthrough, and a six-month exploration phase, then split into a 36-month Phase A with no advisor versus advisor available, and rejoin at the 41-item structural battery followed by the debrief; Study 2 carries no manipulation check.}
\label{fig:s2flow}
\end{figure}
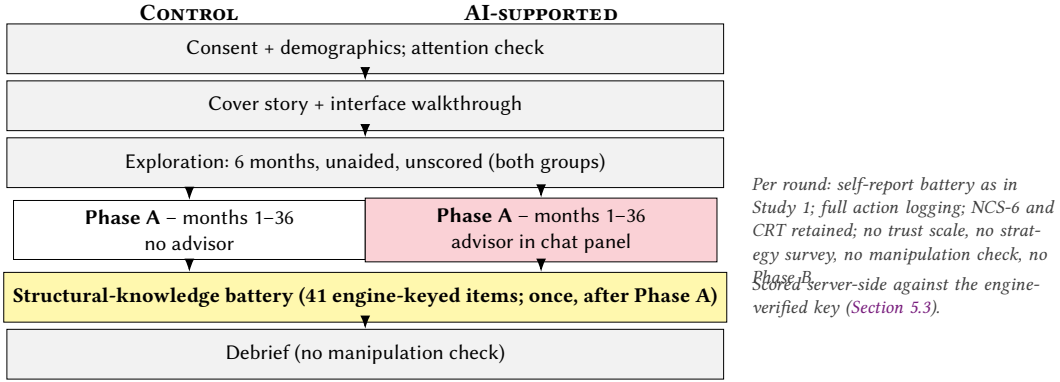

\paragraph{Bayesian estimation and model specifications.} All models use \texttt{brms} \cite{burkner2017brms} over Stan~\cite{carpenter2017stan} in R~4.5.0~\cite{Rcite} (\texttt{brms}~2.23.0, Stan~2.32.2 via \texttt{rstan}~2.32.7, \texttt{lme4}~1.1.37, \texttt{lmerTest}~3.1.3, \texttt{posterior}~1.6.1) \footnote{For guidance, see~\cite{schad2021toward,van2021bayesian,dix2022bayesian,burkner2017brms,kay2016researcher}.}. Chains target $\geq 1{,}000$ effective samples; Gelman--Rubin~\cite{gelman1992inference} must be $< 1.1$. Posteriors: mean and $95\%$ HDI; $p_b$ \cite{shi2020reconnecting} at $p_b \leq 2.5\%$ for zero-distinguishability. \condnoai{} and Phase~A are reference levels. Priors and full specifications in \autoref{app:method-details}.

\emph{Specifications.} (i)~Study~1 equivalence: beta-binomial, $k \,|\, \text{trials}(n) \sim \text{condition}$, two chains. (ii)~Item-response (Study~1 reanalysis; Study~2 primary): 2PL, $\text{logit}\,P(\text{correct}) = \alpha_i\,\theta_p + \delta_i$, $\text{SD}(\theta) = 1$; discriminations \emph{unconstrained} in sign; four chains; Study~2 adds condition on $\theta$. Study~1 reanalysis ($24$ items, $n = 200$): priors $\mathcal{N}(1, 1)$ (discrimination) and $\mathcal{N}(0, 2)$ (difficulty); item-level discrimination and difficulty SDs have half-normal priors with scales $1$ and $2$, respectively, and person SD is fixed at $1$; $6{,}500$ iterations, $1{,}500$ warmup, \texttt{adapt\_delta} $= 0.97$.

\emph{Diagnostics.} Sign reflection per the specification note (\autoref{app:deviations}). \emph{After alignment}: $\hat{R} \leq 1.019$ (items), $1.001$ (persons); bulk-ESS $\geq 419$ (items), $27{,}338$ (persons); tail-ESS $\geq 6{,}566$ and $12{,}297$; no divergent transitions in $20{,}000$ draws. The only parameter short of the $1{,}000$-sample bulk-ESS target is the population discrimination intercept the sign anchoring acts on; every individual item clears it by a wide margin (bulk-ESS $\geq 4{,}803$). Its mixing is bounded by the reflection symmetry rather than by chain length: raising the run from $14{,}000$ to $20{,}000$ draws left $\hat{R}$ unchanged and lowered its bulk-ESS. We report this rather than absorb it; no conclusion in the paper rests on that parameter. The identification could be handled at the likelihood level instead, by bounding discriminations positive, which would remove the reflection mode and the mixing problem with it. We do not, for two reasons: the registered specification leaves discriminations unconstrained in sign, and the negatively discriminating items are a substantive outcome here rather than a nuisance: they are the items the advisor model itself gets wrong (\autoref{sec:limitations}), which a positivity constraint would render unobservable by construction. The likelihood is reflection-symmetric, but the $\mathcal{N}(1,1)$ discrimination prior is not; alignment therefore does not follow from posterior symmetry alone. Per-chain and per-draw orientation agree on all $20{,}000$ retained draws of each condition fit. The reported estimates retain the registered unconstrained model, and their sensitivity to orientation is documented in the reproduction notes. Reliability of $\theta$ is $.765$ on the \emph{full} $24$-item set. Of $24$ items, $7$ have negative discrimination and $9$ fall below $|a| = 0.5$; median $0.637$, range $-0.65$ to $1.89$.

\paragraph{Post-task instruments.} The \condai{} group also completed the 12-item Jian, Bisantz, and Drury Trust in Automated Systems scale~\cite{jian2000foundations} with two custom AI-reliance items (unreported, see \autoref{sec:measures}). The manipulation check asked both groups whether an AI advisor was available (Yes/No); those answering yes also reported consultation frequency (Never / 1--3 / 4--10 / more than 10 times). The strategy survey posed four open-ended prompts: overall strategy; aspects understood well versus unclear; how strategy changed and what triggered changes; and a group-adapted advisor prompt (\condai{}: how suggestions influenced decisions; \condnoai{}: how access to one would have changed their approach), followed by per-variable relevance ratings (1--5) for each of the 10 derived Tailorshop variables. The debrief disclosed the study goal with a free-text suspicion probe.

\paragraph{Prediction-task item structure.} Items span the marketing, production, and human-resources subsystems; several are dynamically counter-intuitive, so a purely structural reading fails. Items were split, condition-blind, into \emph{intuition-aligned} (pooled majority answer matches the engine key) and \emph{intuition-misaligned} halves; an advisor that \emph{teaches} counter-intuitive contingencies would raise \condai{} on the misaligned half, assistance that \emph{erodes} intuitive engagement would lower it on the aligned half; the key was unbalanced ($9$ \emph{higher}, $15$ \emph{lower}).

\paragraph{LLM advisor technical parameters.} Decoding settings are given in \autoref{sec:apparatus}; the system prompt ran to about $1{,}000$ tokens. At the start of Phase~B the \condai{} transition screen carried the exact notice ``The assistant is unavailable for the remaining 12 months of the simulation. Please continue managing the company on your own.'', and the panel was hidden for the duration of Phase~B.

\begin{figure}[!htbp]
\centering
\resizebox{\linewidth}{!}{%
\begin{tikzpicture}[
 font=\sffamily\footnotesize,
 box/.style={draw, rounded corners=2pt, align=center, inner sep=5pt, minimum height=14mm},
 arr/.style={->, >=Latex, semithick},
 lbl/.style={font=\sffamily\scriptsize, align=center, inner sep=2pt, fill=white},
]
\node[box, fill=gray!8, text width=3.6cm] (p) at (0,0)
 {\textbf{Participant}\\dashboard: observable state\\twelve decision controls};
\node[box, fill=cpsai!30, text width=4.2cm] (llm) at (8.5,0)
 {\textbf{LLM advisor} (\texttt{gpt-5.5})\\fixed system prompt\\reasoning \texttt{none}\\sees only the observable state};
\node[box, fill=gray!25, text width=3.9cm] (eng) at (0,-4.5)
 {\textbf{Tailorshop engine}\\server-side; hidden parameters\\identical seed in both groups};
\node[box, fill=orange!30, font=\sffamily\scriptsize,
 text width=4.2cm, minimum height=7mm] at (8.5,-1.6)
 {Phase B: panel hidden\\advisor withdrawn};

\draw[arr] ($(p.east)+(0,.45)$) -- ($(llm.west)+(0,.45)$);
\node[lbl, text width=3.9cm, anchor=south] at (4.1,.55)
 {\textbf{(1) Optional free-text query}\\server adds observable state, four-month trend, and full Phase-A transcript};

\draw[arr] ($(llm.west)+(0,-.45)$) -- ($(p.east)+(0,-.45)$);
\node[lbl, text width=3.9cm, anchor=north] at (4.1,-.55)
 {\textbf{(2) Parsed recommendation}\\per-control values + one-line rationale\\fixed output schema\\\emph{Apply} pre-fills controls};

\draw[arr] ($(p.south)+(-1.5,0)$) -- ($(eng.north)+(-1.5,0)$);
\node[lbl, text width=2.65cm, anchor=west] at (-1.3,-2.05)
 {\textbf{(3) Submit month}\\advice applied or overridden};

\draw[arr] ($(eng.north)+(1.5,0)$) -- ($(p.south)+(1.5,0)$);
\node[lbl, text width=2.65cm, anchor=east] at (1.3,-3.1)
 {\textbf{(4) Next-month state}\\returned to the dashboard};

\draw[dashed, semithick, gray] (llm.east) -- ++(.55,0) |- (eng.east);
\node[fill=white, inner sep=1pt, text=gray, font=\large] at (7,-4.5) {$\times$};
\node[lbl, text=gray, anchor=south] at (7,-4.35) {No actuation rights};
\end{tikzpicture}%
}
\caption{\textbf{The advisor could inform every decision but execute none: it saw only what the participant saw, and its advice reached the simulation only through the participant's own submission.} The \condai{} consultation loop. (1)~The participant may consult the advisor with a free-text query; the server automatically adds the observable dashboard state, a four-month trend summary (company value, bank-account cash, shirt sales and worker satisfaction), and the running Phase-A transcript (the model retains the full transcript; the panel shows only the latest exchange). (2)~The advisor's free-text reply is parsed server-side into a structured recommendation: one value per decision control, plus a one-line rationale, against the output schema fixed in the system prompt; the parsed values render as the panel recommendation and per-control chips, and \emph{Apply} pre-fills the controls with them. (3)~Only the participant submits the month, applied, edited, or ignored advice alike, so the advisor has no actuation rights on the engine (dashed). (4)~The engine returns the next month's state to the dashboard. In Phase~B the panel is hidden and the loop reduces to steps (3)--(4), identical to \condnoai{}.}
\Description{A diagram with three boxes: Participant (dashboard), LLM advisor, and Tailorshop engine. Arrows show an optional query from participant to advisor with server-added context, a recommendation returned to the participant, the participant submitting monthly decisions to the engine, and the engine returning the next state. A dashed crossed-out line between advisor and engine marks that the advisor has no actuation rights. An orange tag notes that in Phase B the advisor panel is hidden. By scrolling, participants in the \condai{} group could see the advisor's full analysis, the options it weighed, and the risks it identified.}
\label{fig:advisor}
\end{figure}
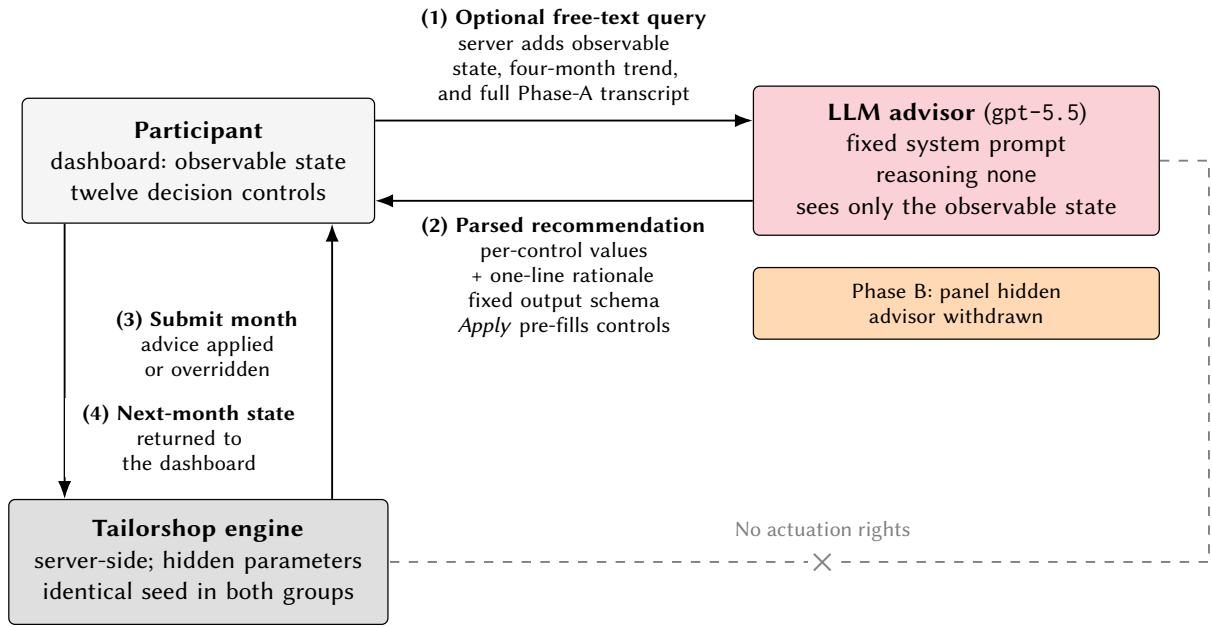

\begin{table}[t]
\caption{\textbf{The advisor's help flowed to the participants least able to survive its
withdrawal.} Stratification on the pre-manipulation exploration-phase baseline, which precedes
any divergence between groups. Values are end-of-phase company value in \$k unless noted;
``weak'' and ``strong'' are the within-group halves of the baseline distribution. All exploratory
and uncorrected.}
\Description{A table stratifying both conditions by pre-manipulation baseline skill, showing
that the baseline gradient predicting performance is flattened under assistance, that the
weak-baseline aided participants lose the most at withdrawal while their unaided counterparts
gain, and that weak-baseline aided participants nonetheless do not end below unaided ones.}
\label{tab:stratification}
\footnotesize
\setlength{\tabcolsep}{4pt}
\begin{tabular}{@{}p{0.34\linewidth}rrl@{}}
\toprule
& \textbf{\condnoai{}} & \textbf{\condai{}} & \textbf{Test} \\
\midrule
\multicolumn{4}{@{}l}{\textit{Baseline gradient on Phase-A outcome}}\\*
Slope per baseline \emph{SD} & $+72.4$k & flattened & $t(196) = 7.41$, $p < .001$ \\
Within-group $r$ & $+.50$ & $+.23$ & interaction $-62.4$k/\emph{SD}, $t(196) = -4.92$, $p < .001$ \\
\addlinespace
\multicolumn{4}{@{}l}{\textit{A$\to$B change in end value, by baseline half}}\\*
Weak half & $+52$k & $-61$k & interaction $+61.0$k/\emph{SD}, \\
Strong half & $+30$k & $-19$k & \quad $t(196) = 4.36$, $p < .001$ \\
\condnoai{} slope alone & \multicolumn{2}{c}{$-38.2$k/\emph{SD}} & $t(196) = -3.54$, $p < .001$ \\
Cognitive-reflection split & \multicolumn{2}{c}{$+6.3$k/\emph{SD}} & $t(196) = 0.44$, $p = .66$, n.s. \\
\addlinespace
\multicolumn{4}{@{}l}{\textit{Tails of the distribution}}\\*
Below do-nothing in \emph{both} phases & $10$ & $1$ & of $11$ such participants \\
\quad months changing nothing & $0.18$ & $0.09$ & \\
Unaided above the worst AI-alone run & $16$ & $5$ & Fisher's exact, $p = .019$ \\
\addlinespace
\multicolumn{4}{@{}l}{\textit{Weak-baseline participants, unaided (Phase~B)}}\\*
Registered do-nothing scale & $0.785$ & $0.744$ & $p = .43$ \\
Ceiling-free persistence scale & $0.373$ & $0.440$ & $p = .073$ \\
\addlinespace
\multicolumn{4}{@{}l}{\textit{Within \condai{}}}\\*
Self-reported reliance $\times$ Phase-B rate & n/a & $r = -.19$ & $p = .063$ \\
\bottomrule
\end{tabular}
\end{table}

\subsection{Supporting robustness checks}\label{app:robustness-details}

\emph{Study~1 battery.} Participants answered \emph{lower} on $54.2\%$ of items against a $62.5\%$ lower-key share, giving a response-bias expectation of $.5105$ versus observed accuracy $.5038$. The eight items with engine effects below $5$k yielded accuracy $.447$ ($95\%$ CI $[.421,.473]$). The other sixteen yielded $.532$. Restricting H1 to those stronger-effect items did not reveal a group difference ($p=.53$).

\emph{Study~1 withdrawal interaction.} Alternative estimates (95\% CIs) were $-1.35$ $[-1.67,-1.02]$ for rising-months scoring, $-0.53$ $[-0.77,-0.29]$ for persistence scoring, and $-1.22$ $[-1.79,-0.65]$ with participant phase slopes. Restricting to firms always above zero in Phase~A gave $-1.18$ $[-1.53,-0.83]$ ($n=184$). Always above $50$k gave $-1.07$ $[-1.44,-0.70]$ ($n=173$). All $p<.001$.

\emph{Study~2 knowledge--performance moderation.} Among 147 solvent participants, the interaction was $+0.14$ ($z=0.50$, $p=.62$). Facet checks use \textsc{iia}, the only adequately reliable facet. Baseline adjustment gave $-0.80$ ($z=-2.93$, $p=.003$). The \textsc{iia}-only check gave $-0.59$ ($p=.070$). Solvent-only selection reduced performance SD from $.31$ to $.23$ in \condnoai{} and $.29$ to $.28$ in \condai{}. Thorndike Case~II correction to each group's original spread left both within-group correlations at $+.13$.

\emph{Study~2 score checks.} Every facet exceeded its chance level by more than a standard deviation. A fixed semantic-option-code response within each facet scores $16/41=.390$, exceeded by $103/198$ participants ($52\%$, binomial $p=.619$ against one-half). This is a specified strategy benchmark, not a general ceiling on knowledge-free responding. Accuracy and condition-centered 2PL scores gave similar performance correlations. Excluding 28 participants with zero rising-months rates left $r=+.182$ ($p=.018$), with rank correlation $\rho=+.235$.

\emph{Study~2 prior sensitivity.} Exploratory four-chain refits (6,000 retained draws each) yielded $+0.42$ (95\% CrI $[+0.12,+0.72]$) with the original prior, $+0.43$ ($[+0.12,+0.74]$) with a symmetric $\mathcal{N}(0,1)$ population-discrimination prior, and $+0.42$ ($[+0.11,+0.75]$) with a $\mathcal{N}(1,0.5^2)$ prior restricted to positive values. Individual item discriminations remained unconstrained. The unrestricted tighter-prior fit mixed poorly ($\hat{R}=1.12$ for the population intercept) and was not counted as robustness evidence. These checks do not establish measurement validity.

\FloatBarrier
\section{Bayesian evidence for the null claims}\label{app:bayes-nulls}
Every null the argument rests on carries a Bayesian counterpart here (the confirmatory contrasts of both studies, the mental-model tests, and the calibration correlations) so a reader can separate the three cases a bare $p > .05$ conflates: show \emph{for} the null, show for an effect too small to matter, and simple lack of power. Nothing in these tables is hand-transcribed; each row is emitted directly from the per-family scripts (\texttt{audit\_deep/pe\_bayes\_*.R}), which recompute the frequentist statistic and fit the Bayesian counterpart in the same pass; exploratory families are reported in the materials release and are not repeated here. Two-group contrasts use a default Cauchy prior on the standardized effect ($r = 0.707$); correlations the corresponding prior; contingency tables an independent-multinomial sampling model; multilevel and item-response nulls are estimated in \texttt{brms} via Savage--Dickey density ratio. $\mathrm{BF}_{01}$ above $3$ is moderate evidence for the null; above $10$ is strong. The ``in ROPE'' column gives the posterior mass inside $\pm 0.40$ standardized units, the same region used for the confirmatory tests, and carrying the same caveat: \emph{practically equivalent} means no effect as large as $d = 0.40$, not the absence of smaller effects.

Historical corrections are documented in the audit scripts. The pooled calibration row was removed because it treated correlations from the same participants as independent; separate correlations remain. The unreproducible Phase-A dropout log-rank claim was also withdrawn.

\emph{The discrimination-sign split.} These rows classify each of the $41$ Study~2 items on the sign of its posterior mean discrimination, a threshold at zero that takes no account of whether that sign is resolved. Often it is not: $19$ of the $41$ items have a $95\%$ credible interval for $a$ spanning zero, and the three nearest the cut sit at $P(a < 0) = .49$, $.48$ and $.59$, so which side they fall on is settled by sampling noise rather than by the data. Restricting the split to the $22$ items whose interval excludes zero sharpens the interaction rather than removing it: $\chi^2(1) = 12.82$, $p = .0003$, $d = +0.53$ ($95\%$ CrI $[+0.24, +0.82]$), against $\chi^2(1) = 4.52$, $p = .034$, $d = +0.17$ ($[+0.02, +0.32]$) on the full set. This selected-item analysis is exploratory: the same responses determine item inclusion and estimate the contrast. Its direction is consistent with a difference in $\theta$: the group scoring higher on positively discriminating items should score lower on negatively discriminating ones. The table reports the full split, which is the registered form. The sharpened estimate rests on only two negatively discriminating items and is offered as a sensitivity, not a replacement.

\begin{table*}[tbp]
\centering\footnotesize\setlength{\tabcolsep}{3.5pt}
\caption{\textbf{Confirmatory nulls, both studies.} $\mathrm{BF}_{01}$ is the Bayes factor for the null over the alternative, so larger values favour \emph{absence} of the effect; \emph{moderate} marks $\mathrm{BF}_{01} \geq 3$, \emph{strong} $\geq 10$, and \emph{inconclusive} $1/3 < \mathrm{BF}_{01} < 3$---which is \emph{not} support for the null. $^{\dagger}$ marks a value between $3$ and $3.5$, i.e. only just past the moderate threshold. The effect column is the posterior median with its $95\,\%$ credible interval, on the standardised $d$ scale (or $\rho$ for correlations and latent $\theta$ for the 2PL); ``in ROPE'' is the share of the posterior inside the $\pm 0.40$ region of practical equivalence used throughout the paper. \emph{Non-zero, but inside the ROPE} marks the one case the other labels cannot express: a posterior whose credible interval excludes zero while lying almost entirely within $\pm 0.40$, so the effect is directional and within the registered, relatively broad ROPE. That is a different finding from evidence of no effect, and is not counted as one. BF labels concern the point null; ROPE labels concern magnitude. A positive interval need not settle either question. Frequentist $p$ values refer to the specified tests.}
\Description{Bayesian counterparts for the registered confirmatory contrasts. These carry the H1 equivalence verdict and the Phase-B level check.}
\label{tab:bayesnull_conf}
\par\smallskip\noindent\textit{$^{\ddagger}$Direction positive; practical magnitude unresolved; no BF reported.}\par\smallskip
\begin{tabular}{p{0.34\textwidth}rrrlrl}
\toprule
Measure & $n$ & $p$ & $\mathrm{BF}_{01}$ & Effect [95\,\% CrI] & in ROPE & Evidence \\
\midrule
per-item prediction accuracy (24 items) & 200 & .379 & --- & $+0.04$ [$-0.05$, $+0.12$] & 100\,\% & in ROPE \\
per-participant mean accuracy & 200 & .384 & 4.6 & $+0.12$ [$-0.16$, $+0.39$] & 98\,\% & moderate \\
month-1 beat indicator vs do-nothing reference & 200 & .642 & 5.6 & $+0.08$ [$-0.26$, $+0.41$] & 97\,\% & moderate \\
month-2 beat indicator vs do-nothing reference & 200 & .875 & 6.2 & $+0.03$ [$-0.31$, $+0.37$] & 98\,\% & moderate \\
state-conditional beat rate, months 2-12 Phase B & 200 & .749 & 6.5 & $-0.05$ [$-0.30$, $+0.20$] & 100\,\% & moderate \\
per-participant Phase-B beat rate (do-nothing reference) & 200 & --- & 6.5 & $-0.01$ [$-0.28$, $+0.26$] & 100\,\% & moderate \\
per-participant battery accuracy (41 items) & 198 & .241 & 3.4 & $+0.15$ [$-0.12$, $+0.43$] & 96\,\% & moderate$^{\dagger}$ \\
condition coefficient on theta, brms 2PL, sign-aligned draws & 198 & --- & --- & $+0.42$ [$+0.11$, $+0.73$] & 46\,\% & positive$^{\ddagger}$ \\
item-level correctness, condition x facet interaction (2 df) & 198 & .531 & 38 & $+0.01$ [$-0.14$, $+0.16$] & 100\,\% & strong \\
item-level correctness, condition x discrimination-sign interaction (1 df) & 198 & .034 & 0.5 & $+0.17$ [$+0.02$, $+0.32$] & 100\,\% & \emph{non-zero, but inside the ROPE} \\
item-level correctness, condition x discrimination-sign interaction (1 df) & 200 & .048 & 0.9 & $+0.15$ [$+0.01$, $+0.30$] & 100\,\% & \emph{non-zero, but inside the ROPE} \\
\bottomrule
\end{tabular}
\end{table*}

\begin{table*}[tbp]
\centering\footnotesize\setlength{\tabcolsep}{3.5pt}
\caption{\textbf{Calibration nulls: self-rating against measured knowledge.} $\mathrm{BF}_{01}$ is the Bayes factor for the null over the alternative, so larger values favour \emph{absence} of the effect; \emph{moderate} marks $\mathrm{BF}_{01} \geq 3$, \emph{strong} $\geq 10$, and \emph{inconclusive} $1/3 < \mathrm{BF}_{01} < 3$---which is \emph{not} support for the null. $^{\dagger}$ marks a value between $3$ and $3.5$, i.e. only just past the moderate threshold. The effect column is the posterior median with its $95\,\%$ credible interval, on the standardised $d$ scale (or $\rho$ for correlations and latent $\theta$ for the 2PL); ``in ROPE'' is the share of the posterior inside the $\pm 0.40$ region of practical equivalence used throughout the paper. \emph{Non-zero, but inside the ROPE} marks the one case the other labels cannot express: a posterior whose credible interval excludes zero while lying almost entirely within $\pm 0.40$, so the effect is directional and within the registered, relatively broad ROPE. That is a different finding from evidence of no effect, and is not counted as one. BF labels concern the point null; ROPE labels concern magnitude. A positive interval need not settle either question. Frequentist $p$ values refer to the specified tests.}
\Description{Bayesian counterparts for the self-rating-versus-accuracy correlations in both studies. These quantify associations between self-ratings and measured knowledge.}
\label{tab:bayesnull_subj}
\begin{tabular}{p{0.34\textwidth}rrrlrl}
\toprule
Measure & $n$ & $p$ & $\mathrm{BF}_{01}$ & Effect [95\,\% CrI] & in ROPE & Evidence \\
\midrule
mean self-rated understanding (1-7) vs prediction-battery accuracy & 100 & .996 & 6.5 & $-0.00$ [$-0.19$, $+0.19$] & --- & moderate \\
mean self-rated understanding (1-7) vs prediction-battery accuracy & 100 & .253 & 3.4 & $+0.11$ [$-0.08$, $+0.30$] & --- & moderate$^{\dagger}$ \\
mean per-round decision confidence (0-100 VAS) vs prediction-battery accuracy & 100 & .743 & 6.2 & $-0.03$ [$-0.23$, $+0.16$] & --- & moderate \\
mean per-round decision confidence (0-100 VAS) vs prediction-battery accuracy & 100 & .733 & 6.2 & $+0.03$ [$-0.16$, $+0.23$] & --- & moderate \\
mean self-rated understanding (1-7) vs prediction-battery accuracy & 200 & .355 & 6.0 & $+0.06$ [$-0.07$, $+0.20$] & --- & moderate \\
mean per-round decision confidence (0-100 VAS) vs prediction-battery accuracy & 200 & .924 & 9.2 & $+0.01$ [$-0.13$, $+0.15$] & --- & moderate \\
delta Fisher-z of r(understanding, accuracy) between groups & 200 & .421 & 5.0 & $+0.11$ [$-0.17$, $+0.39$] & --- & moderate \\
delta Fisher-z of r(confidence, accuracy) between groups & 200 & .636 & 6.0 & $+0.07$ [$-0.21$, $+0.34$] & --- & moderate \\
mean understanding (months 1-36) x battery prop. correct & 101 & .102 & 1.8 & $+0.16$ [$-0.04$, $+0.34$] & --- & \emph{inconclusive} \\
mean understanding (months 1-36) x battery prop. correct & 97 & .117 & 1.9 & $+0.15$ [$-0.04$, $+0.34$] & --- & \emph{inconclusive} \\
mean confidence (months 1-36) x battery prop. correct & 101 & .705 & 6.1 & $+0.04$ [$-0.16$, $+0.22$] & --- & moderate \\
mean confidence (months 1-36) x battery prop. correct & 97 & .311 & 3.9 & $+0.10$ [$-0.09$, $+0.29$] & --- & moderate \\
total controls moved/month x battery prop. correct & 198 & .667 & 8.4 & $-0.03$ [$-0.17$, $+0.11$] & --- & moderate \\
total controls moved/month x battery prop. correct & 101 & .617 & 5.8 & $-0.05$ [$-0.24$, $+0.15$] & --- & moderate \\
total controls moved/month x battery prop. correct & 97 & .697 & 6.0 & $-0.04$ [$-0.24$, $+0.16$] & --- & moderate \\
\bottomrule
\end{tabular}
\end{table*}

\begin{table*}[tbp]
\centering\footnotesize\setlength{\tabcolsep}{3.5pt}
\caption{\textbf{Mental-model nulls.} $\mathrm{BF}_{01}$ is the Bayes factor for the null over the alternative, so larger values favour \emph{absence} of the effect; \emph{moderate} marks $\mathrm{BF}_{01} \geq 3$, \emph{strong} $\geq 10$, and \emph{inconclusive} $1/3 < \mathrm{BF}_{01} < 3$---which is \emph{not} support for the null. $^{\dagger}$ marks a value between $3$ and $3.5$, i.e. only just past the moderate threshold. The effect column is the posterior median with its $95\,\%$ credible interval, on the standardised $d$ scale (or $\rho$ for correlations and latent $\theta$ for the 2PL); ``in ROPE'' is the share of the posterior inside the $\pm 0.40$ region of practical equivalence used throughout the paper. \emph{Non-zero, but inside the ROPE} marks the one case the other labels cannot express: a posterior whose credible interval excludes zero while lying almost entirely within $\pm 0.40$, so the effect is directional and within the registered, relatively broad ROPE. That is a different finding from evidence of no effect, and is not counted as one. BF labels concern the point null; ROPE labels concern magnitude. A positive interval need not settle either question. Frequentist $p$ values refer to the specified tests.}
\Description{Bayesian counterparts for the Study 1 mental-model contrasts that stand behind the H1 verdict.}
\label{tab:bayesnull_model}
\begin{tabular}{p{0.34\textwidth}rrrlrl}
\toprule
Measure & $n$ & $p$ & $\mathrm{BF}_{01}$ & Effect [95\,\% CrI] & in ROPE & Evidence \\
\midrule
metacog\_sens (conf correct - conf incorrect) & 200 & .428 & 4.8 & $+0.11$ [$-0.17$, $+0.37$] & --- & moderate \\
auroc2 (type-2 AUROC, confidence resolution) & 200 & .286 & 3.8 & $+0.14$ [$-0.13$, $+0.41$] & --- & moderate \\
declarative importance rating spread (within-participant SD of 10 ratings) & 200 & .903 & 6.5 & $+0.02$ [$-0.25$, $+0.28$] & --- & moderate \\
Phase-B signed company-value forecast error (\$k) & 196 & .797 & 6.4 & $-0.02$ [$-0.29$, $+0.25$] & --- & moderate \\
beat (B2-B4 first unaided months) & 600 & .499 & 2.5 & $-0.11$ [$-0.47$, $+0.25$] & 94\,\% & \emph{inconclusive} \\
beat | prior unaided month missed & 430 & .589 & 2.8 & $+0.08$ [$-0.27$, $+0.43$] & 96\,\% & \emph{inconclusive} \\
\bottomrule
\end{tabular}
\end{table*}

\FloatBarrier
\section{Prompts used in the study}
\label{app:prompts}
All large-language-model prompts used in the deployed study ship byte-exact with the materials
release (\texttt{prompts/}): the advisor system prompt, the per-query context template, the
decision-extraction prompt and the AI-alone benchmark prompts. The advisor prompt states the
objective, lists the observable state and the twelve controls with their ranges and step sizes,
fixes the reply structure (analysis, options, recommendation, risks, and a JSON block carrying all
twelve decisions), and forbids any claim about engine internals; the horizon token
\texttt{\{max\_months\}} is substituted at call time ($12$ in Study~1, $36$ in Study~2). No prompt
contains the simulation's formulas or coefficients: the advisor and the AI-alone benchmark know
only what a participant can observe in the interface.

\end{document}